\documentclass[a4paper, amsfonts, amssymb, amsmath, reprint,prl, superscriptaddress,showkeys, nofootinbib,twoside,twocolumn]{revtex4-2}

\usepackage{parskip}

\usepackage[english]{babel}
\usepackage[utf8]{inputenc}
\usepackage{bbold,dsfont}
\usepackage[mathscr]{eucal}
\usepackage{braket}

\usepackage{comment}
\usepackage{subfigure}
\usepackage[colorinlistoftodos, color=green!40, prependcaption]{todonotes}
\usepackage{lineno}

\usepackage[pdftex, pdftitle={Article}, pdfauthor={Author}]{hyperref} 
\begin{document}
\title{Quantum versatility in PageRank}

\author{Wei-Wei Zhang$^{\dagger}$}
\affiliation{%
School of Computer Science, Northwestern Polytechnical University, Xi’an 710129, China 
}
\author{Zheping Wu$^{\dagger}$}

\affiliation{%
School of Computer Science, Northwestern Polytechnical University, Xi’an 710129, China 
}
\author{Hengyue Jia}
\affiliation{School of Information, Central University of Finance and Economics, Beijing 100081, China}

\author{Wei Zhao}
\email{zhaowei9801@163.com}
\affiliation{
National Key Laboratory of Security Communication,  Chengdu 610041, China 
}

\author{Qingbing Ji}
\email{jqbxy@163.com}
\affiliation{
National Key Laboratory of Security Communication,  Chengdu 610041, China 
}

\author{Wei Pan}

\affiliation{%
School of Computer Science, Northwestern Polytechnical University, Xi’an 710129, China 
}

\author{Haobin Shi}
\email{shihaobin@nwpu.edu.cn}
\affiliation{%
School of Computer Science, Northwestern Polytechnical University, Xi’an 710129, China 
}

\date{\today} 

\begin{abstract}
Quantum mechanics empowers the emergence of quantum advantages in various fields, including quantum algorithms. Quantum PageRank is a promising tool for a future quantum internet. Recently, arbitrary phase rotations (APR) have been introduced in the underlying Szegedy’s quantum walk of 
quantum PageRank algorithm. In this work, we thoroughly study the role APR plays in quantum PageRank.  We discover the versatility resulting from quantumness.  Specifically, we discover the emergence of a cluster phenomenon in rankings considering the rotation phases, i.e. the existence of similar clusters in the distribution of the rankings and their fidelity with the corresponding classical PageRanks, the ranking distribution variance, the coherence and entanglement of PageRank states,  and the power law parameter in the ranking distributions on a scale-free network concerning the two rotation phases. Furthermore, we propose an alternate quantum PageRank with APR which provides an extra tunnel for the analysis of PageRank. We also study the PageRank on the trackback graph of a scale-free graph for the investigation of network information traffic tracking. We demonstrate the rich cluster diversity formed in our alternate quantum PageRank, which offers a novel perspective on the quantum versatility of PageRank. Our results present the quantum-enabled perspective for PageRanking and shed light on the design and application of practical quantum PageRank algorithms.
\end{abstract}


\maketitle

\section{Introduction} \label{sec:introduction}

 With the development of quantum and artificial intelligence technology, the emergence of quantum advantages in various scenarios~\cite{Arute2019,Zhong2020,Zhu2022,Spagnolo2023,He2024DeRL,Li2023Using,Li2023Using}  enhances people's expectations of the quantum era. One of the main developments in quantum technology is quantum communication and quantum internet, which has the 
potential to revolutionize the way we communicate and exchange information~\cite{kimble2008quantum,wehner2018quantum,RevModPhys.95.045006,chen2021integrated,brito2020statistical}. Compared to the challenge in the development of a universal quantum computer, 
the quantum internet requires fewer resources and is expected to become available 
before having a fault-tolerant quantum computer. 

Data mining is an important field for the development of the Internet. Google PageRank is one of the most influential algorithms and is the most objective and efficient one~\cite{page1999pagerank,paparo2012google}. 
High-quality ranking
of a large number of web pages in Google search engine is ensured with PageRank
algorithm, which has been 
identified as one of the top 10 algorithms in data mining~\cite{wu2008top},
and one of the nine algorithms that changed the future~\cite{maccormick2013nine}. 
With the development of the quantum internet, the search algorithms in quantum have attracted massive attention. By considering the importance of quantum superposition and entanglement, the first 
PageRank algorithm in quantum internet was proposed~\cite{paparo2012google}, where search progress was based on  Szegedy quantum walks~\cite{szegedy2004quantum}. With the consideration of various complex graphs and quantum evolution operators, quantum search algorithm and PageRank demonstrates its advantage over its classical counterpart~\cite{grover1996fast,paparo2013quantum,chawla2020discrete,sanchez2012quantum,loke2017comparing,tang2021tensorflow,wang2020experimental,wang2014enhanced, grover1997quantum,  long1999phase,  galindo2000family, 
 long2002phase,  li2007phase,  toyama2008multiphase}.

Recently, a generalized quantum PageRank introduced arbitrary phase rotations (APR) in the underlying
Szegedy’s quantum walk and defined three different APR schemes, where  
the authors conjecture that the opposite-phases  algorithm seems to be very promising for scale-free graphs,
 since it resembles the classical distribution of PageRanks
 highlighting truly secondary hubs, and maintaining the quantum  stability~\cite{PhysRevResearch.5.013061}.In this work, we thoroughly study the role APR plays in quantum PageRank and discover the versatility cluster phase phenomenon 
resulting from quantumness. Specifically, we discover the emergence of a cluster phase phenomenon in
rankings considering the rotation phases, while considering the distribution of the
rankings and their fidelity with the corresponding classical PageRank, the ranking distribution variance, the coherence and entanglement in the PageRank states, and the power law parameter in the ranking distributions on a scale-free network.  We find the clusters formed in these quantities are strongly correlated with each other. 
We propose an alternate quantum PageRank model that offers a rich Pagerank cluster diversity for data mining in the quantum internet, where the correlations between the clusters formed by variance, quantum fidelity, coherence, entanglement, and the $\beta$ power show different shapes and correlations. 
The PageRank distributions from APR model and the alternate APR models, both emphasize the same set of nodes, which makes our results convincing.  We discover that the probability of each node generated with different PageRank models demonstrates differences, which enables our PageRank model to offer a multi-perspective interpretation of the network structure and suits the study of complex network scenarios. 
Furthermore, we study PageRank on the trackback graph of a scale-free graph and discover that PageRank demonstrates a different shape of clusters and emphasizes different sets of nodes, which demonstrate the important nodes in the network information traffic tracking.  
Our results present
the quantum versatility phases for PageRanking in different scenarios and offer novel perspectives for the quantum data mining method. Our work 
sheds light on the design and application of 
practical quantum PageRank algorithms.

\section{The PageRank with arbitrary phase rotations} 
\label{sec:the model}
In this section, we introduce the generalized quantum PageRank with arbitrary phase rotations (APR)~\cite{PhysRevResearch.5.013061}, where  
the Google matrix $\mathbf{G}$ used for PageRank is defined using transition matrix $\mathbf{E}$ and the all ones matrix $\mathbf{1}$ the directed graph as follows,
\begin{equation}
    \mathbf{G}=\alpha\mathbf{E}+\frac{1-\alpha}{N}\mathbf{1}
    \label{eq:G}
\end{equation}
with $\mathbf{E}$ defined as 
\begin{equation}
E_{i,j}:=
\left\{\begin{array}{c}1/\mathrm{outdeg}(P_j)\quad\mathrm{if~}P_j\in B_i,\\
1/N\quad\mathrm{if~outdeg}(P_j)=0,\\0\quad\mathrm{otherwise},\end{array}
\right.
\end{equation}where outdeg($P_j$) is the outdegree of node $P_j$ and $B_i$ is the set of nodes liking to the node $P_i$. We set $\alpha=0.85$ in this study.

The  PageRank vector $I_c$ should satisfy $I_c = \mathbf{G}I_c$, which means 
$I_c$ is the eigenvector of the matrix $\mathbf{G}$ with eigenvalue 1. The dynamical evolution of random walks mimics the power law which ensures a random walk
performed with the matrix $\mathbf{G}$ over any probability distribution
will converge to the eigenvector $I_c$.  

The generalized quantum PageRank on a directed graph is defined based on Szegedy quantum walks~\cite{szegedy2004quantum} using the Google matrix defined in Eq.~\ref{eq:G}~\cite{PhysRevResearch.5.013061}.
The basis of the evolution is the $N\times N$ directed edges of the duplicated graph $\{\ket{i}_1\ket{j}_2, i, j\in N\times N\}$ with the indexes 1 and 2 referring to the nodes on two copies of the original graph. In our study, we set the initial state as the equal probability
distribution state as follows,
\begin{equation}
\ket{\psi_0}=\frac{1}{\sqrt{N}}\sum_{i=1}^N\ket{\psi_i},
\end{equation}
with $\ket{\psi_i}:=\ket{i}_1\otimes\sum_{k=1}^N\sqrt{G_{ki}}\ket{{k}}_2$.

\begin{figure*}[t]
    \centering
{{\subfigure(a)}\includegraphics[width=0.5\columnwidth]{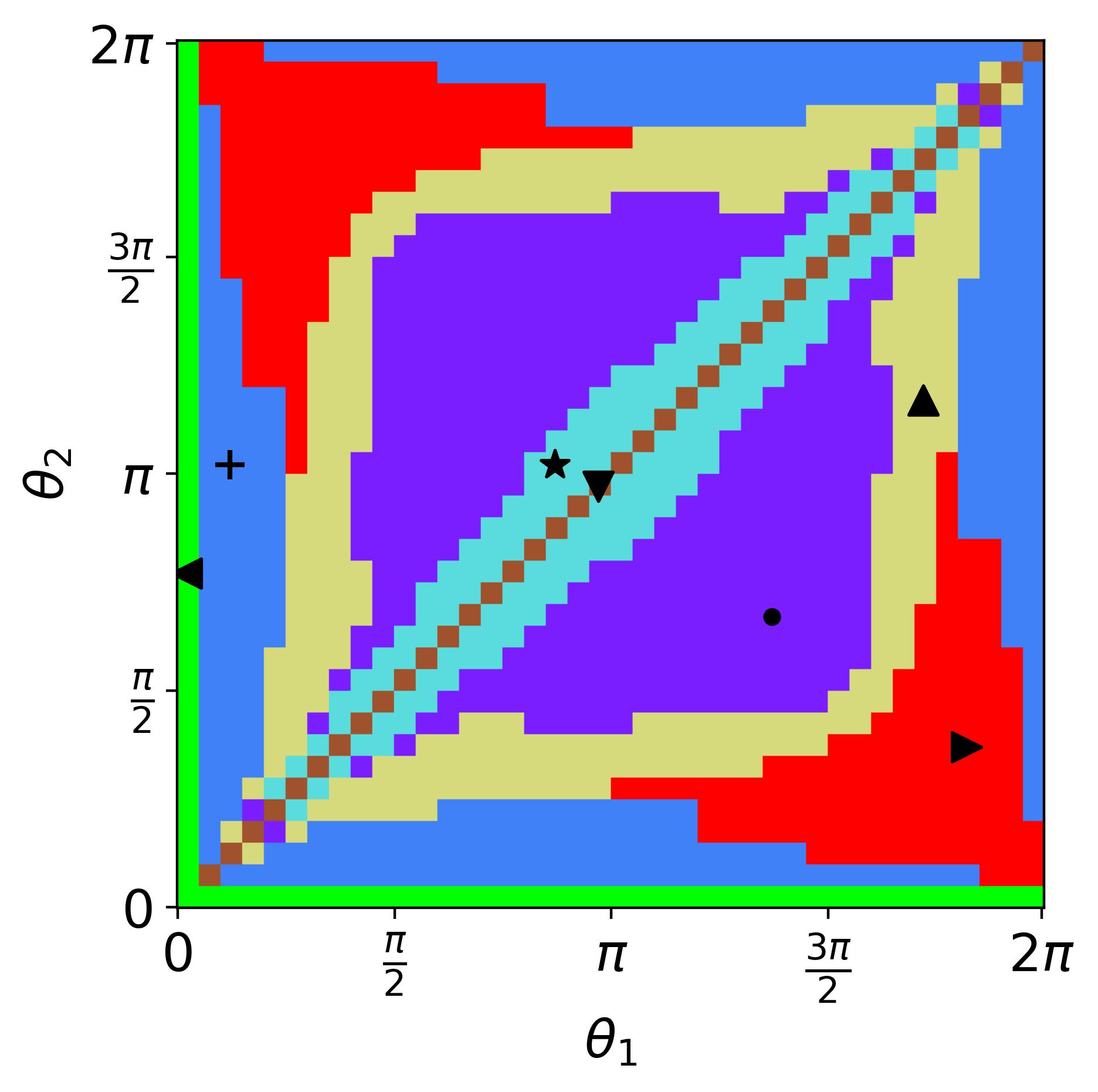}}
{{\subfigure(b)}
\includegraphics[width=0.6\columnwidth]{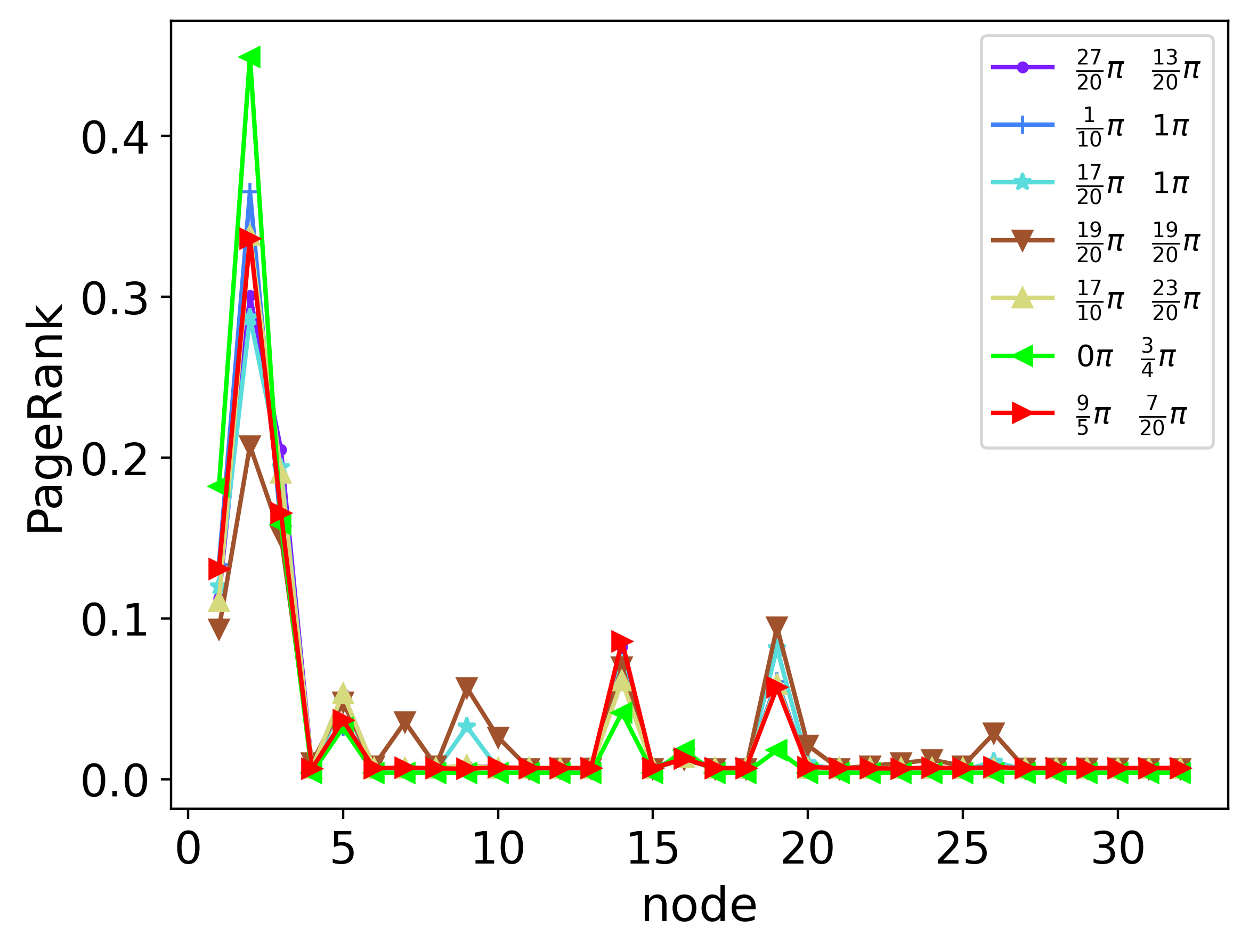}}
{{\subfigure(c)}
\includegraphics[width=0.7\columnwidth]{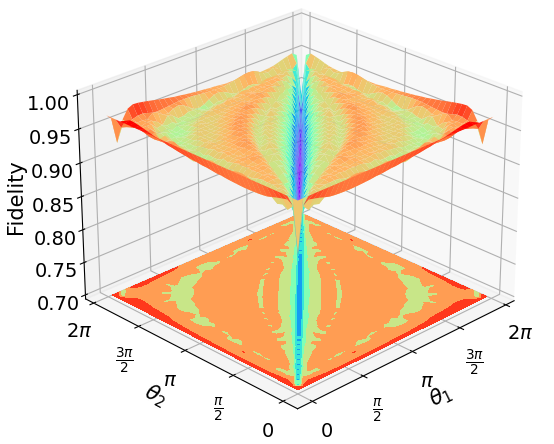}}
    \caption{(a) The clusters of PageRank  distributions $\protect\overrightarrow{\bf{I}}$  with respect to $\theta_1$ and $\theta_2$, the color encodes the different cluster areas, (b) the comparisons of the typical PageRank distributions in each cluster found in (a), the corresponding $(\theta_1, \theta_2)$ parameters are noted with same markers in (a), (c) The distribution of quantum PageRank fidelity  between quantum PageRank $\Vec{I}$ and the classical PageRank in $(\theta_1, \theta_2)$ space. The color represents the value of the corresponding Fidelity value.}
    \label{fig:knn-typical distribution}
\end{figure*}

The quantum PageRank  evolution is governed by the unitary operator defined as follows
\begin{equation}
 U(\theta):=S([1-e^{i\theta}]\Pi-\mathds{1})  
\end{equation}
with $\theta\in [0,2\pi)$, $\mathds{1}$ as the identity operator and 
\begin{align}
    \Pi:&=\sum_{k=1}^N\ket{\psi_k}\bra{\psi_k}\\
   S:&=\sum_{i,j}\ket{i,j}\bra{j,i}.
\end{align}
Since the swap operator $S$ changes the directedness of the graph, the operator $U$ has to be applied an even number of times to ensure the consistency of the graph, i.e. the evolution operator is set as 
\begin{equation}
W(\theta_1,\theta_2):=U(\theta_2)U(\theta_1)
\label{eq:W}
\end{equation}
The evolution state of quantum PageRank is 
\begin{equation}
\ket{\psi_f(t)}=W(\theta_1,\theta_2)^t\ket{\psi_0}.
\end{equation} 
The PageRank outcomes are represented by probabilities of the nodes on the 2nd copy of the graph which is obtained by quantum projection 
\begin{equation}
    \label{eq:Iqt}    I_q(P_i,t):=||\bra{i}\psi_f(t)\rangle||_2^2,
\end{equation}
which is named instantaneous PageRank and fluctuates in time~\cite{PhysRevResearch.5.013061}.
\begin{figure*}[ht]
    \centering
{
    {\subfigure(a)}\includegraphics[width=0.76\columnwidth]{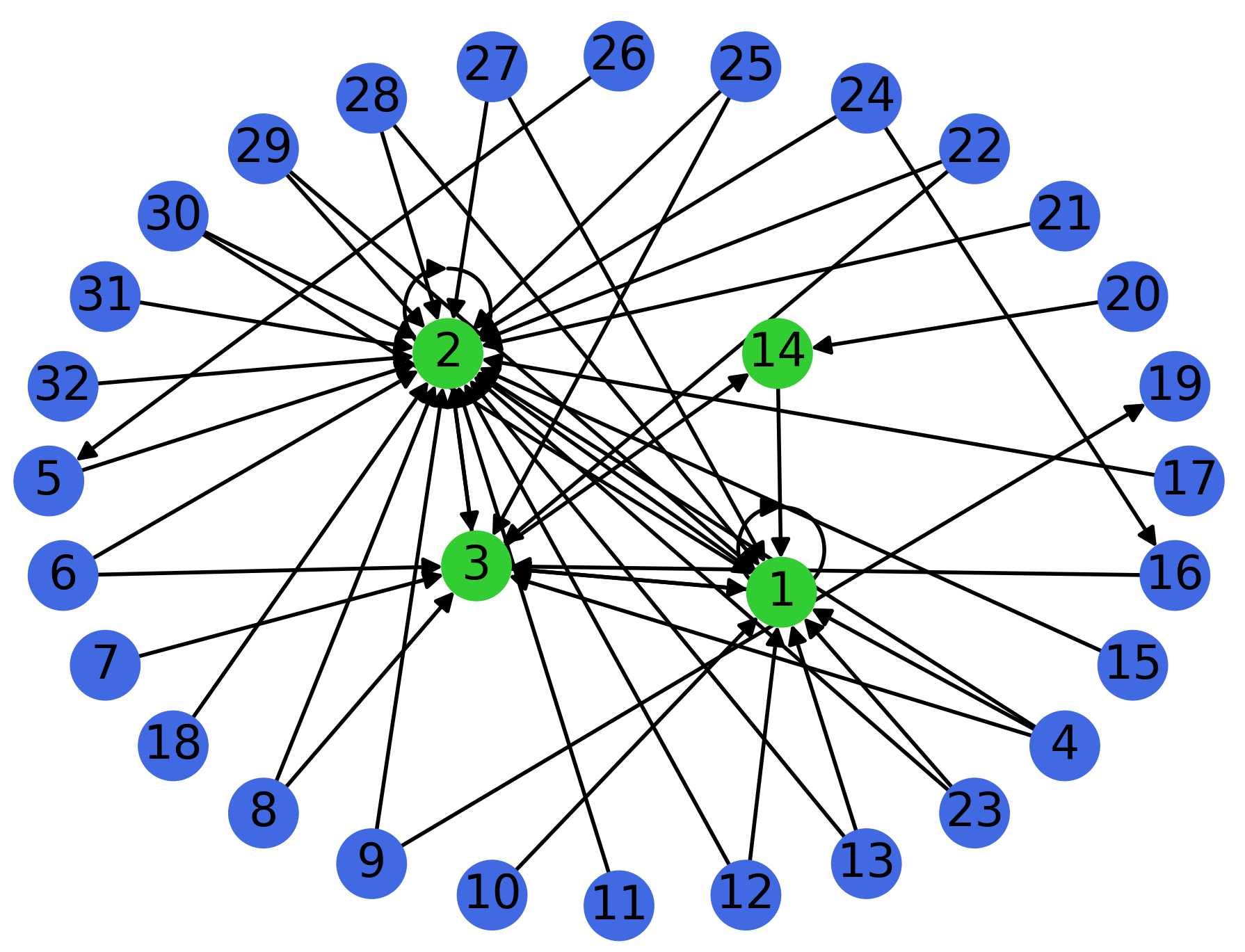}
    {
        {\subfigure(b)}\includegraphics[width=0.76\columnwidth]{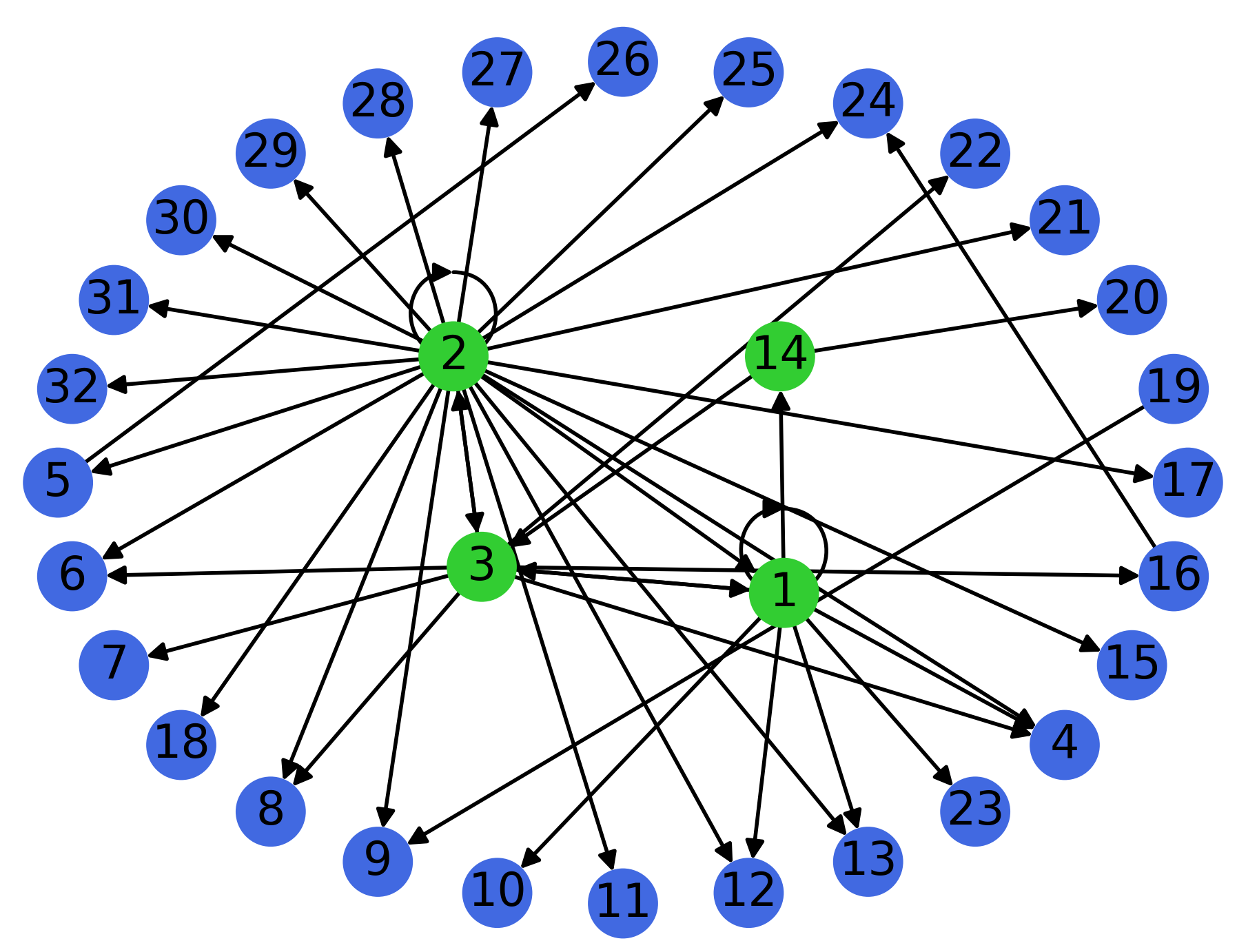}
        \caption{(a) The 32 nodes scale-free graph used in our simulations, (b) the trackback graph of the 32 nodes scale-free graph in (a).}
        \label{fig:graph}
    }
}
\end{figure*}

\section{Cluster phases in quantum PageRank with APR}
\label{sec:clusters-PageRank}
In this section, we study the signatures of corresponding PageRank on scale-free graphs concerning the adjustable rotations parameters.  Specifically, the cluster phases we discovered in the PageRank outcomes, the ranking distribution variances, the  quantum PageRank fidelity, the coherence and entanglement in the quantum PageRank states and the power law parameter with respect to the two adjustable phase rotations.

Compared with Ref.~\cite{PhysRevResearch.5.013061}, where the authors study three specified cases of quantum PageRank with APR, equal-phases, opposite-phases, and alternate-phases, and find the quantumness do play an important role, In this work, we thoroughly study the whole $(\theta_1,\theta_2)$ APR region, and characterize the PageRank behavior.

\textbf{PageRank distributions} quantify the importance of the nodes in a directed graph. The more important a linking node is, the higher its PageRank probability is. Since the instantaneous PageRank in Eq.~\ref{eq:Iqt} fluctuates in time, we define a time-averaged quantum PageRank as following

\begin{align}
    I_q(P_i):=\frac{1}{\Delta t}\sum_{t=T-\Delta t}^TI_q(P_i,t),
\end{align}

where we set $\Delta t=500$ and $T=5000$, which ensures our modified PageRank $I_q(P_i)$ evaluates the averaged value as the probabilities oscillate stably. Note that our definition with the sum starting from $T-\Delta t$ is different but equivalent to the traditional definition starting from 0.  As $t$ gets larger the instantaneous 
 fluctuations of the
PageRank probability gets smaller and more stable, which is part of the data we are
interested in and reflects the intrinsic properties of the quantum PageRank and is operational friendly.

With a set of $(\theta_1,\theta_2)$, we could obtain a PageRank outcomes which is a probability distribution on the node of a directed graph. By scanning through the $\theta_1\in[0,2\pi),\theta_2\in[0,2\pi))$ parameter region, we obtain the corresponding PageRank distributions $\overrightarrow{\bf{I}}=\{\overrightarrow{I}(\theta_1,\theta_2)=\{I_1,I_2,...I_N\}\}$. Using the most traditional data clustering techniques, i.e. KNN~\cite{taunk2019brief}, we discover a cluster phases phenomenon in the obtained PageRank Distributions $\overrightarrow{\bf{I}}$, as shown in Fig.~\ref{fig:knn-typical distribution} (a).  Specifically, we firstly cluster the distributions $\overrightarrow{\bf{I}}$ with a way of unsupervised learning and label all the distributions 
$\{\overrightarrow{I}(\theta_1,\theta_2)\}$ with the clustered tags. Lastly, we return the tags each $\{\overrightarrow{I}(\theta_1,\theta_2)\}$ in the ($\theta_1,\theta_2$) parameter region. By encoding the cluster tags as different colors, we demonstrate the PageRank clusters as shown in Fig.~\ref{fig:knn-typical distribution} (a), where we see the seven clusters of the quantum PageRank with APR  considering their PageRank distributions. Here we experiment with a 32-node scale-free graph as shown in Fig.~\ref{fig:graph}(a). 
The typical representative  PageRank 
in each cluster phase is marked in Fig.~\ref{fig:knn-typical distribution}(a), and the corresponding probability distributions are shown in Fig.~\ref{fig:knn-typical distribution}(b), where we see that for each cluster  
the main hub in our chosen  32-node scale-free graph includes the same 10 nodes (nodes indexed as 1,2,3,5,7,9,10,14,19,26), but 
the PageRank within different cluster phases 
shows different PageRank values for each of these 9 nodes to emphasize their corresponding importance.

\textbf{Quantum PageRank fidelity} is used to quantify the similarity between quantum PageRank distribution and the corresponding classical PageRank distribution in Ref.~\cite{PhysRevResearch.5.013061}, which we also use to investigate our PageRanks. The definition of quantum PageRank fidelity is defined as follows,
\begin{equation}
f(I_1,I_2):=\sum_{i=1}^N\sqrt{I_1(P_i)I_2(P_i)},
\end{equation}
with $I_1$ and $I_2$ as the quantum PageRank distribution and the classical PageRank distribution respectively. In Fig.~\ref{fig:knn-typical distribution} (c), we plot the distribution of the fidelity  between quantum PageRank $\overrightarrow{I}$ and the classical
PageRank in ($\theta_1,\theta_2$) space, where we see a  similar cluster pattern with the KNN clusters in Fig.~\ref{fig:knn-typical distribution}(a).

\begin{figure*}[t]
    \centering
{\subfigure(a)}\includegraphics[width=0.44\columnwidth]{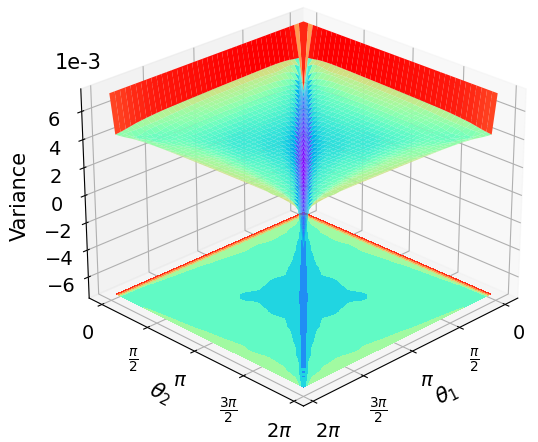}
{\subfigure(b)}\includegraphics[width=0.44\columnwidth]{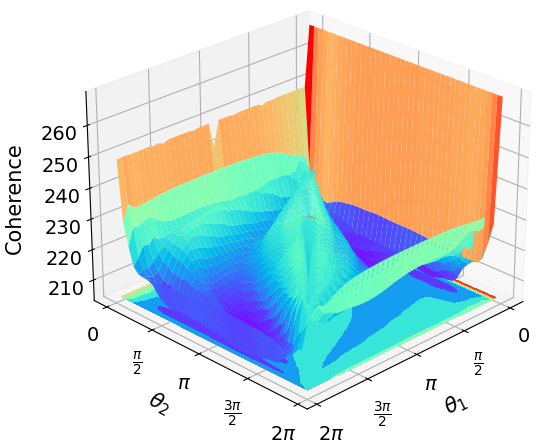}
{\subfigure(c)}\includegraphics[width=0.44\columnwidth]{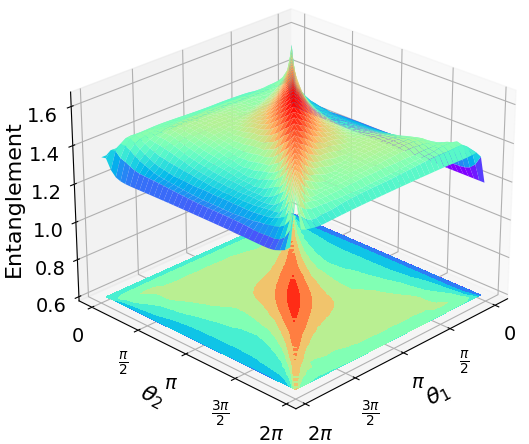}
{\subfigure(d)}\includegraphics[width=0.44\columnwidth]{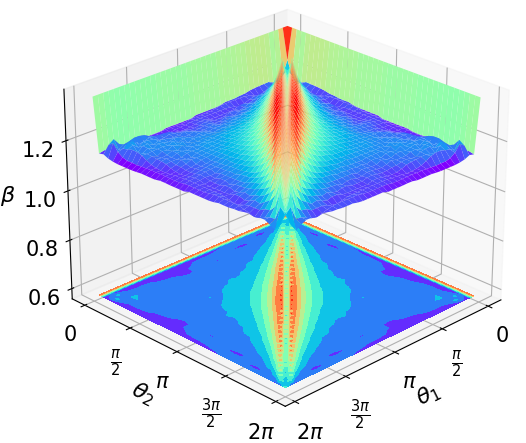}
       \caption{(a) The distribution of the variance of PageRank  distributions $\protect\overrightarrow{\bf{I}}$   with respect to $\theta_1$ and $\theta_2$,(b) the distribution of the coherence and (c) the entanglement of the PageRank quantum state, (d) the distribution of the $\beta$ power of PageRank distributions $\protect\overrightarrow{\bf{I}}$ concerning $\theta_1$ and $\theta_2$. Note: the color represents the value of the y axis quantity.}
    \label{fig:variance-coherence-entanglement-beta}
\end{figure*}

\begin{figure*}[ht]
    \centering
\includegraphics[width=0.8\columnwidth]{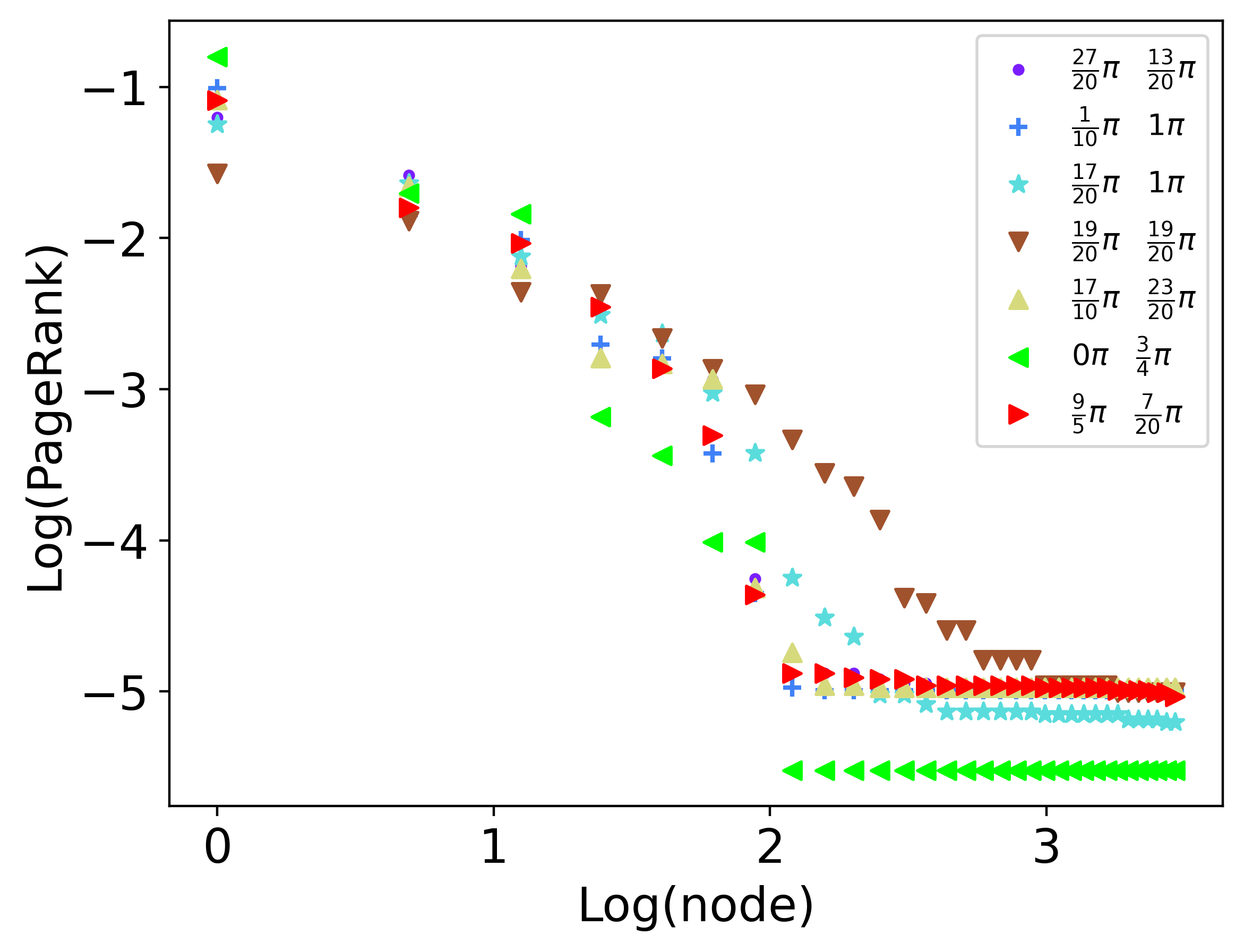}
    \caption{The comparisons of the typical PageRank distribution power-law relations where the $\theta_1,\theta_2$ chosen to generate the data is noted with the same makers in Fig.~\ref{fig:knn-typical distribution}(a), the color of the markers are consistent with the color of the  clusters it belongs to in Fig.~\ref{fig:knn-typical distribution}(a),  the values of $(\theta_1, \theta_2)$ are same as the ones in the legend of Fig.~\ref{fig:knn-typical distribution}(b).}
    \label{fig:beta}
\end{figure*}

\textbf{PageRank distribution variance} $\sigma(\theta_1, \theta_2)$ is the variance of the PageRank distribution $\overrightarrow{I}(\theta_1,\theta_2)$, which quantifies the  dispersion of a probability distribution. In PageRank, the importance of nodes on a directed graph is demonstrated with the probability value of the Ranking outcomes. Therefore, the variance is closely related to the number of important hubs and the number of nodes in each hub. The distribution of the PageRank variance $
\overrightarrow{\sigma}=\{\sigma(\theta_1,\theta_2)\}$ is shown in Fig.~\ref{fig:variance-coherence-entanglement-beta}(a), which indeed shows the cluster phases of PageRanks and the distribution of the emergent cluster phases are consistent with the cluster phases shown in the clustering of PageRank distributions in Fig.~\ref{fig:knn-typical distribution}(a).

\textbf{Coherence and entanglement}, being at the heart of interference
phenomena play a critical role in quantum algorithms as they enable applications that are impossible within classical mechanics. With evolutions of quantum PageRank states, the coherence and entanglement therein also evolve. As a quantum characteristic, the coherence $\mathcal{C}_q$  and entanglement $\mathds{E}_q$ in the PageRank outcome state are important indicators of the quantum resources. 

In this work, we study the distribution of coherence $\mathcal{C}_q$  and entanglement $\mathds{E}_q$ in PageRank outcome states with respect to the adjustable parameters $(\theta_1,\theta_2)$. 
As shown in Fig.~\ref{fig:variance-coherence-entanglement-beta}(b) and (c), there are emergent cluster phases in the distribution of coherence and entanglement, and these cluster phases are consistent with the phases we observed in PageRank distributions and their quantum fidelity with the corresponding classical PageRank in Fig.~\ref{fig:knn-typical distribution}(a, c)  and their variance~Fig.\ref{fig:variance-coherence-entanglement-beta}(a).

For the study of coherence and entanglement in quantum PageRank, we first obtain the coherence and entanglement in the instantaneous quantum PageRank states $\hat{\rho}(t)$ before the measurement. In our case,  we study the reduced density matrix $\hat{\rho}_\text{r}(t)$ obtained by taking the partial trace of the two-copy graph quantum state over one copy of it (we trace over the graph index 2 in our simulations). 
The coherence quantifier for an instantaneous quantum PageRank state $\hat{\rho}_\text{r}(t)$ we used in this work is the intuitive $l_1$-norm of coherence, which is defined~\cite{baumgratz2014quantifying,yue2017bounds} as 
\begin{equation}
    C_{l_1}(\hat{\rho}_\text{r}(t))=\sum_{\substack{i,j \\ i\neq j}}|\rho_{i,j}^\text{r}(t)|.\label{eq:coherence},
\end{equation}
where $\rho_{i,j}^\text{r}$ is the elements in the reduced density matrix of the quantum PageRank state $\hat{\rho}_\text{r}(t)$. 
For a quantum PageRank state with density matrix $\rho(t)$,  the entanglement $\mathcal{E}(\rho_\text{r}(t))$ between the subspace formed by the nodes in each of the two copies of the directed graph is measured by the entropy of the reduced density matrix $\rho_\text{r}(t)$,
\begin{equation}
    \mathcal{E}(\rho_\text{r}(t))=-\lambda_i\text{log}(\lambda_i),
    \label{eq:entanglement}
\end{equation}
where $\lambda_i$ are the eigenvalues of the reduced density matrix $\rho_\text{r}(t)$.
The coherence and entanglement for quantum PageRank is a time-averaged value of the coherence and entanglement in the instantaneous quantum PageRank states. The definitions for quantum 
PageRank coherence $\mathcal{C}_q$ and entanglement $\mathds{E}_q$ are as follows,
\begin{align}
    \mathcal{C}_q&=\frac{1}{\Delta t}\sum_{t=T-\Delta t}^T C_{l_1}(\hat{\rho_\text{r}}(t))\\
    \mathds{E}_q&=\frac{1}{\Delta t}\sum_{t=T-\Delta t}^T\mathcal{E}(\rho_\text{r}(t))
\end{align}

\textbf{Power law distribution} in the connectivity of the nodes is the most deterministic signature of a scale-free graph, the PageRank distributions on a scale-free graph also follow the similar behavior~\cite{PhysRevResearch.5.013061}. The 32-node graph we used in this work as shown in Fig.~\ref{fig:graph}(a) is a scale-free graph. The PageRank can be expressed as 
\begin{equation}
    I_i\sim i^{-\beta}, 
    \label{eq:e-beta}
\end{equation}
where $i$ is the index of the node after sorting them by importance and $\beta$ is a constant coefficient. By taking logarithms to both sides of Eq.~\ref{eq:e-beta}, we obtain 
\begin{equation}
    \text{log}~I_i\sim -\beta~\text{log}i
    \label{eq:beta-function}
\end{equation}
With linear data fitting to the PageRank distributions $\overrightarrow{\bf{I}}$, we could obtain the $\beta(\theta_1,\theta_2)$ values in the whole adjustable parameters region. The distribution of $\beta$ with respect to $(\theta_1,\theta_2)$ is shown in Fig.~\ref{fig:variance-coherence-entanglement-beta}(d), where we see  $\beta$ shows phase phenomena consistency with the cluster phases shown in PageRank outcomes, their quantum fidelity with the corresponding classical PageRank, their variance, the coherence and entanglement distributions in adjustable parameters region. In Fig.~\ref{fig:beta}, we show the logarithmic plot of the PageRanks and the node index (after sorting) for the typical of each cluster.

The degeneracy of the residual nodes of the scale-free graph from the  Logarithmic plot of our quantum PageRanks is closely related with the smoothness of the $\beta$ curve. 
We discover that in all quantum PageRanks clusters, the degeneracy of residual nodes are broken in a certain degree. Specifically, for our standard quantum PageRanks with the case of $\theta_1=\theta_2$ have the largest level of degeneracy broken and have a smooth behavior in the logarithmic plot due to the degeneracy breaking of the less important nodes, which also results its the least fidelity with the classical PageRank. In the other hand, 
  for our standard quantum PageRanks with the case of $\theta_1=-\theta_2$ have restored several degeneracy of the residual nodes compared with the $\theta_1=\theta_2$ case,  and its PageRank logarithmic plot has a more abrupt decay due to the degeneracy restore, which also results in its the higher fidelity with the classical PageRank. 
  These observations are consistent with the results in Rf.~\cite{PhysRevResearch.5.013061}, which are specific cases of our $\theta_1=\theta_2$ and $\theta_1=-\theta_2$ clusters.

\begin{figure*}[ht]
    \centering
{{\subfigure(a)}\includegraphics[width=0.43\columnwidth]{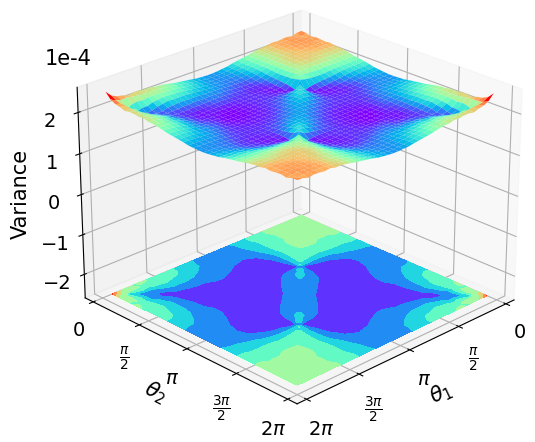}}
{{\subfigure(b)}\includegraphics[width=0.43\columnwidth]{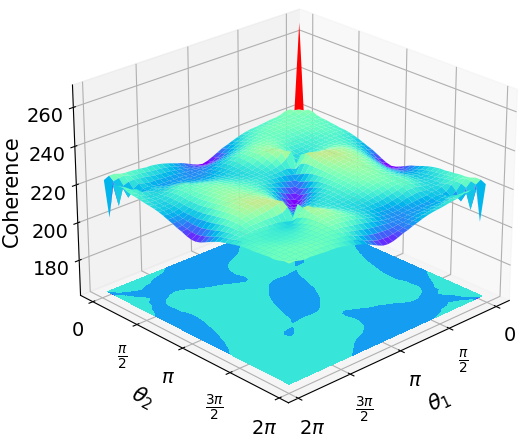}}
{{\subfigure(c)}\includegraphics[width=0.43\columnwidth]{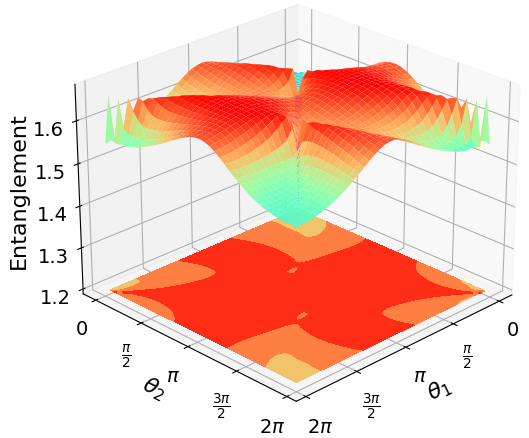}}
{{\subfigure(d)}\includegraphics[width=0.43\columnwidth]{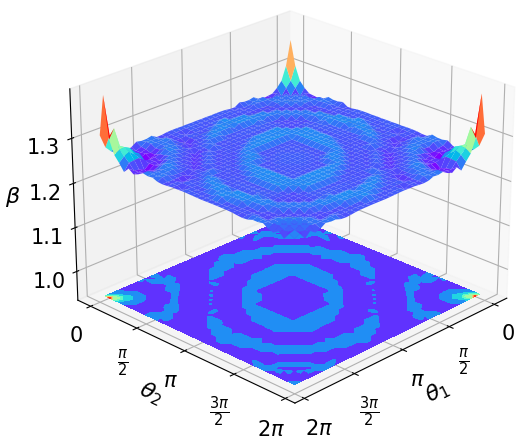}}
    \caption{(a) The distribution of the variance of the alternate equal PageRank distributions $\protect\overrightarrow{\bf{I}}$. (b) the distribution of the alternate equal PageRank state coherence and (c) entanglement (d) the $\beta$ power distribution      
    with respect to $\theta_1$ and $\theta_2$ in alternate equal PageRank. Note: the color represents the value of the y axis quantity.} \label{fig:clusters-alternate}
\end{figure*}

\begin{figure*}[ht]
    \centering
{{\subfigure(a)}\includegraphics[width=0.8\columnwidth]{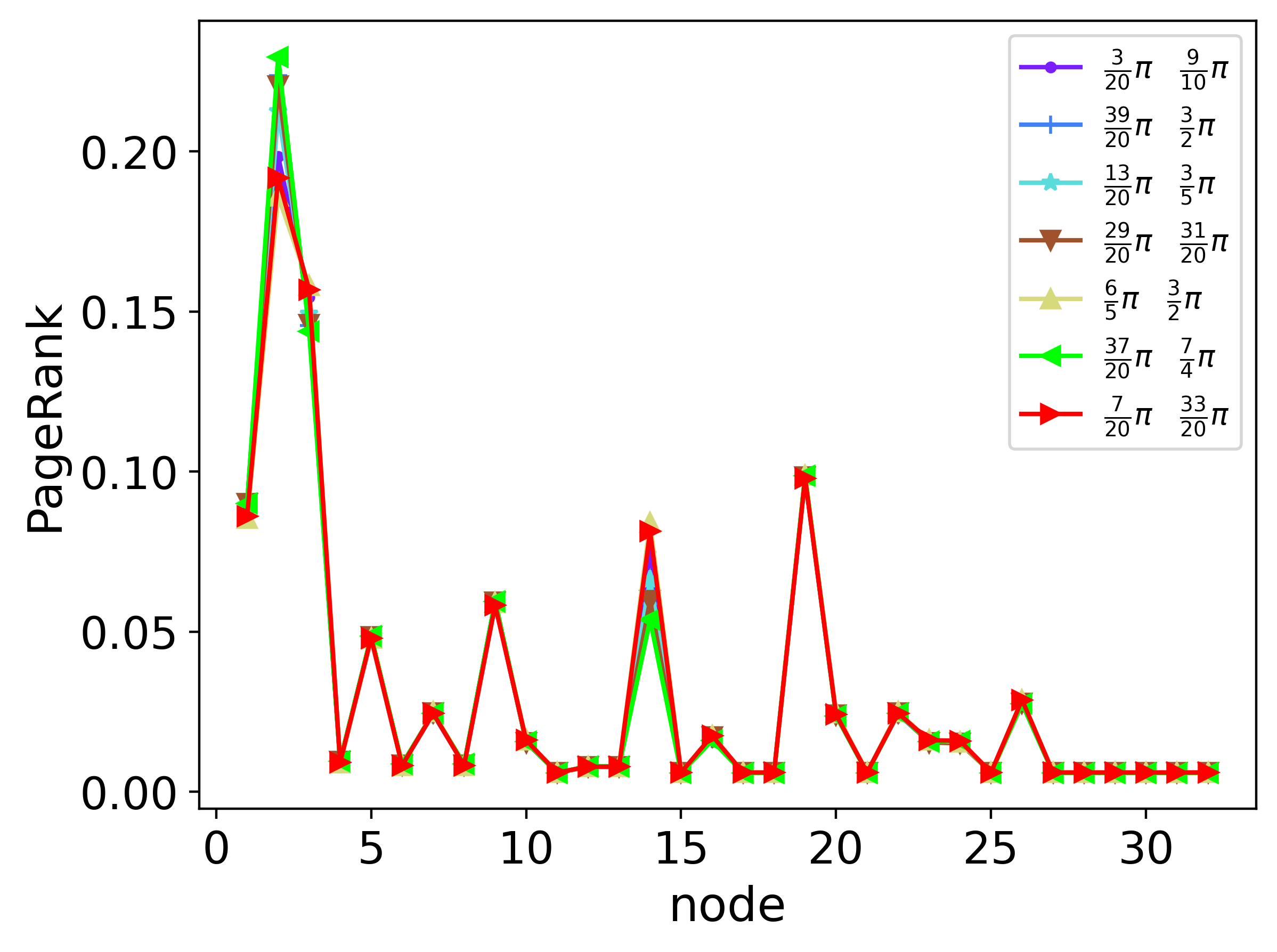}}
{{\subfigure(b)}\includegraphics[width=0.8\columnwidth]{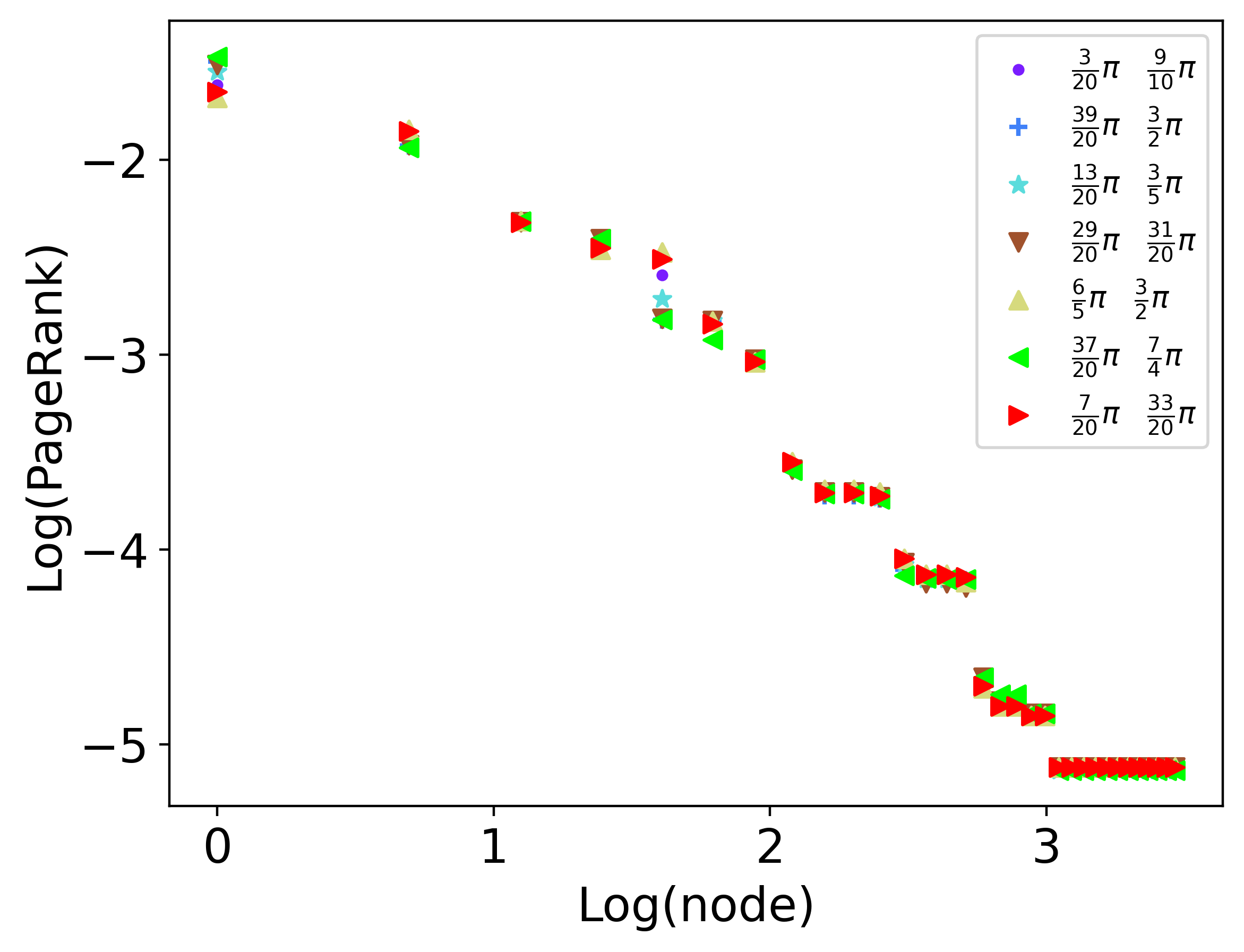}}
    \caption{(a) The comparison of typical PageRank distribution in alternate equal quantum PageRank. (b) the comparison of the typical PageRank distribution power-law relation in alternate equal quantum PageRank. The $\{\theta_1,\theta_2\}$ chosen to generate the data is noted with the same makers in Fig.~\ref{fig:markers-alternate- trackback}(a). Note: the color of the markers in (a) and (b) are consistent with the color of the clusters it belongs
to in Fig.~\ref{fig:markers-alternate- trackback}(a), the values of $(\theta_1, \theta_2)$ are shown in the legend of (a).} \label{fig:clusters-alternate-typical}
\end{figure*}

\textbf{The cluster phases}  in PageRank distributions and their variance, quantum PageRank fidelity, coherence, entanglement, and power law distribution demonstrate similar cluster regions and boundaries, which shows the consistency of these quantities. 
We discover the quantum PageRank fidelity is positively correlated with the variance of the PageRank probability distribution, and anti-correlated with coherence, entanglement, and the $\beta$ coefficient of the quantum PageRank. Our discovery suggests that the quantumness in the schemes is anti-correlated with the variance of the PageRank, and the fidelity with the classical PageRank, but positive-correlated with the coherence, entanglement, and the $\beta$. Specifically,  
 the lower the fidelity with the classical PageRank, the lower the variance of the PageRank probability distribution is, and the higher the coherence, entanglement, and $\beta$ coefficient are.

With a further detailed comparisons of Figs.~\ref{fig:variance-coherence-entanglement-beta} (a-d), we discover a positive correlation between the coherence, entanglement and the $\beta$ parameter of the power-law relation for the PageRank in a scale-free graph. While the variance is anti-correlated to the quantities mentioned above. 
These correlations can be interpreted as follows, (1) the variance is 
a measure of dispersion, i.e. a measure of how far a set of numbers is spread out from their average value. The higher the variance is, the more dispersed the distribution is. The variance quantifies the dispersion of PageRank distribution, which is closely related to the value of diagonal components in the density matrix of PageRank state as shown in Eq.~\ref{eq:Iqt}. (2)  Coherence arising from quantum superposition is a common necessary condition for both entanglement and other types of quantum correlations. 
From the definition of coherence in Eq.~\ref{eq:coherence}, we understand the value of coherence quantifies the absolute values of off-diagonal components in the density matrix of the PageRank state. 
(3) the entanglement as defined in Eq.~\ref{eq:entanglement}, closely related the eigen-spetrum of the reduced density matrix. The higher the entanglement is, the more correlated the nodes in two copies of the graph.
(4) the power law in the PageRank of a scale-free graph as in Eq.~\ref{eq:beta-function} denotes how fast the logarithm of PageRank value drops with the logarithm of node index. The higher the value of $\beta$ is, the faster the PageRank drops, the more dispersed the PageRank is, and the higher the variance of PageRank distribution is. The dominating reason for the cluster phenomenon is the PageRank rule parameters $(\theta_1,\theta_2)$  offer different emphases on the importance of node connections in a network graph.  Therefore, quantum PageRank offers a special versatility for data mining.

\begin{figure*}[!ht]
    \centering
{{\subfigure(a)}\includegraphics[width=0.44\columnwidth]{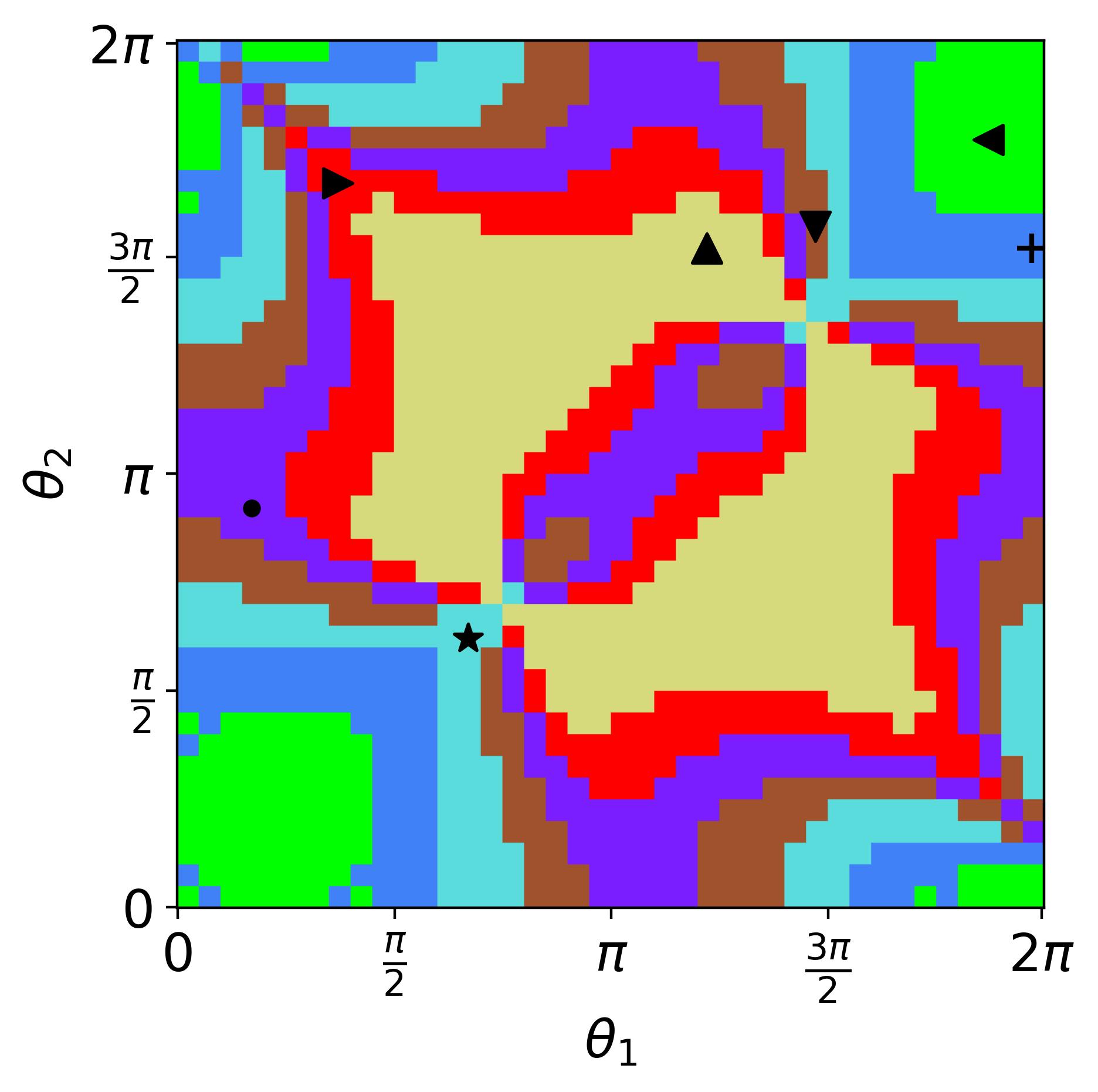}}
{{\subfigure(b)}\includegraphics[width=0.44\columnwidth]{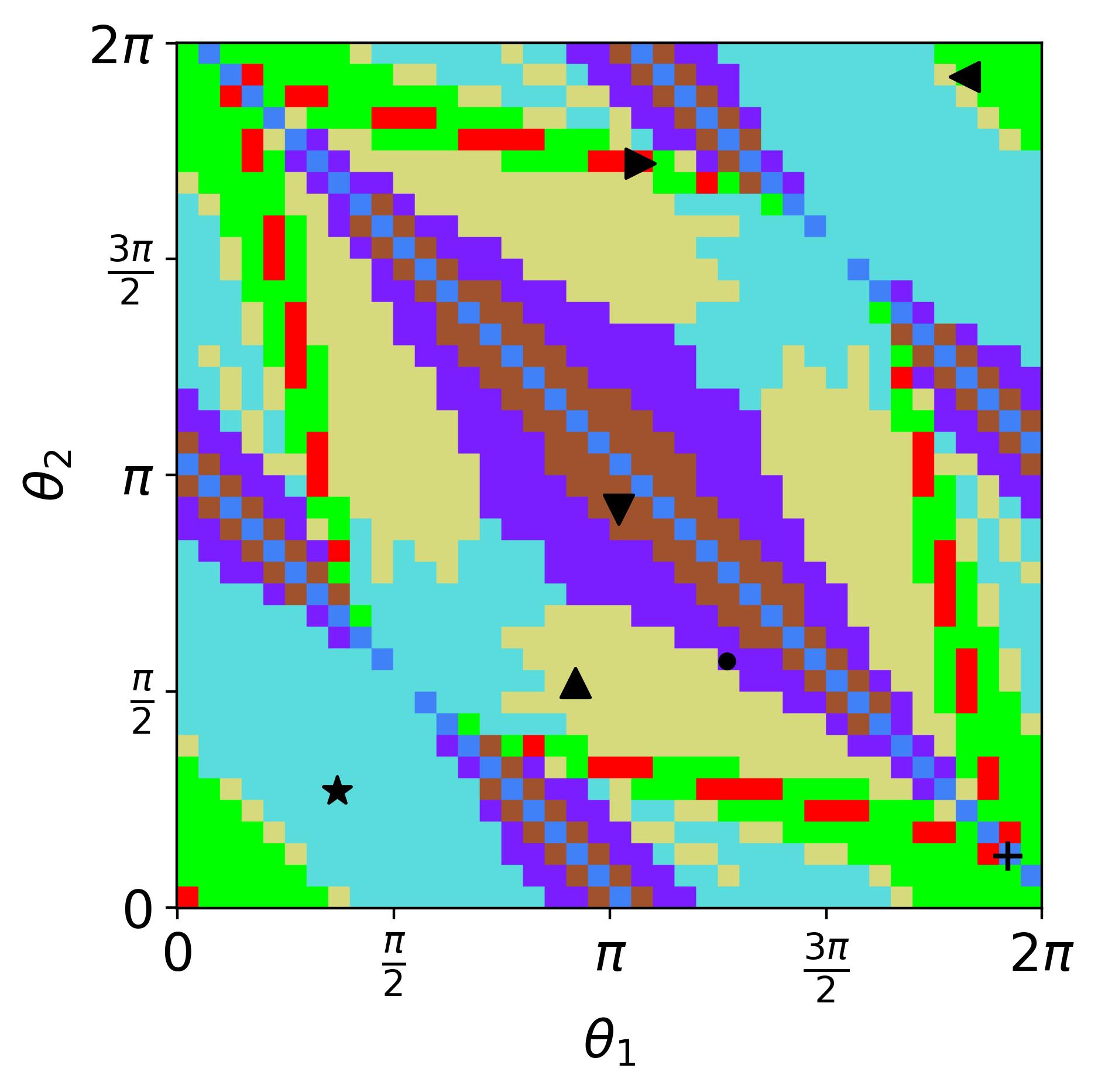}}
{{\subfigure(c)}\includegraphics[width=0.44\columnwidth]{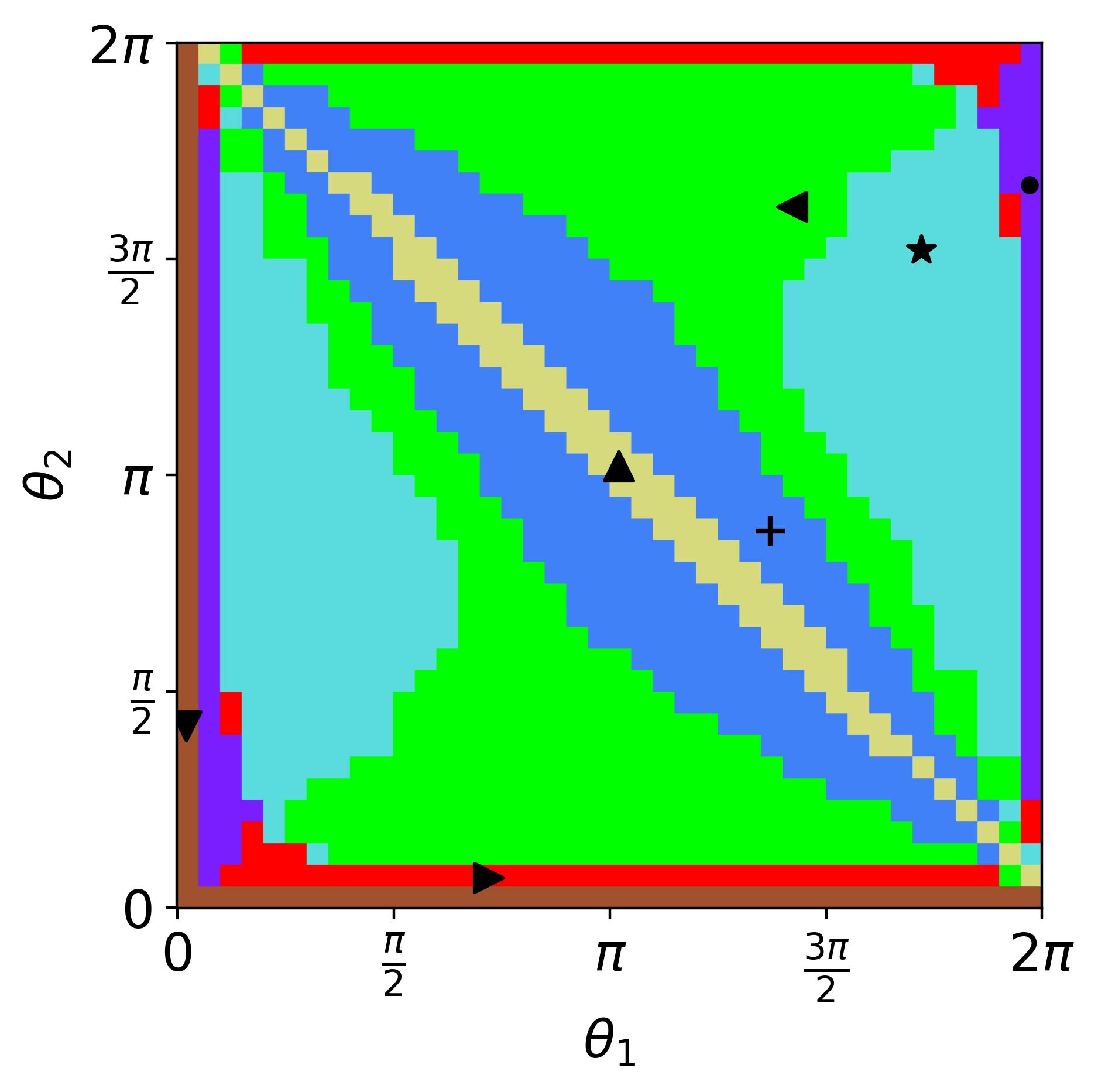}}
{{\subfigure(d)}\includegraphics[width=0.44\columnwidth]{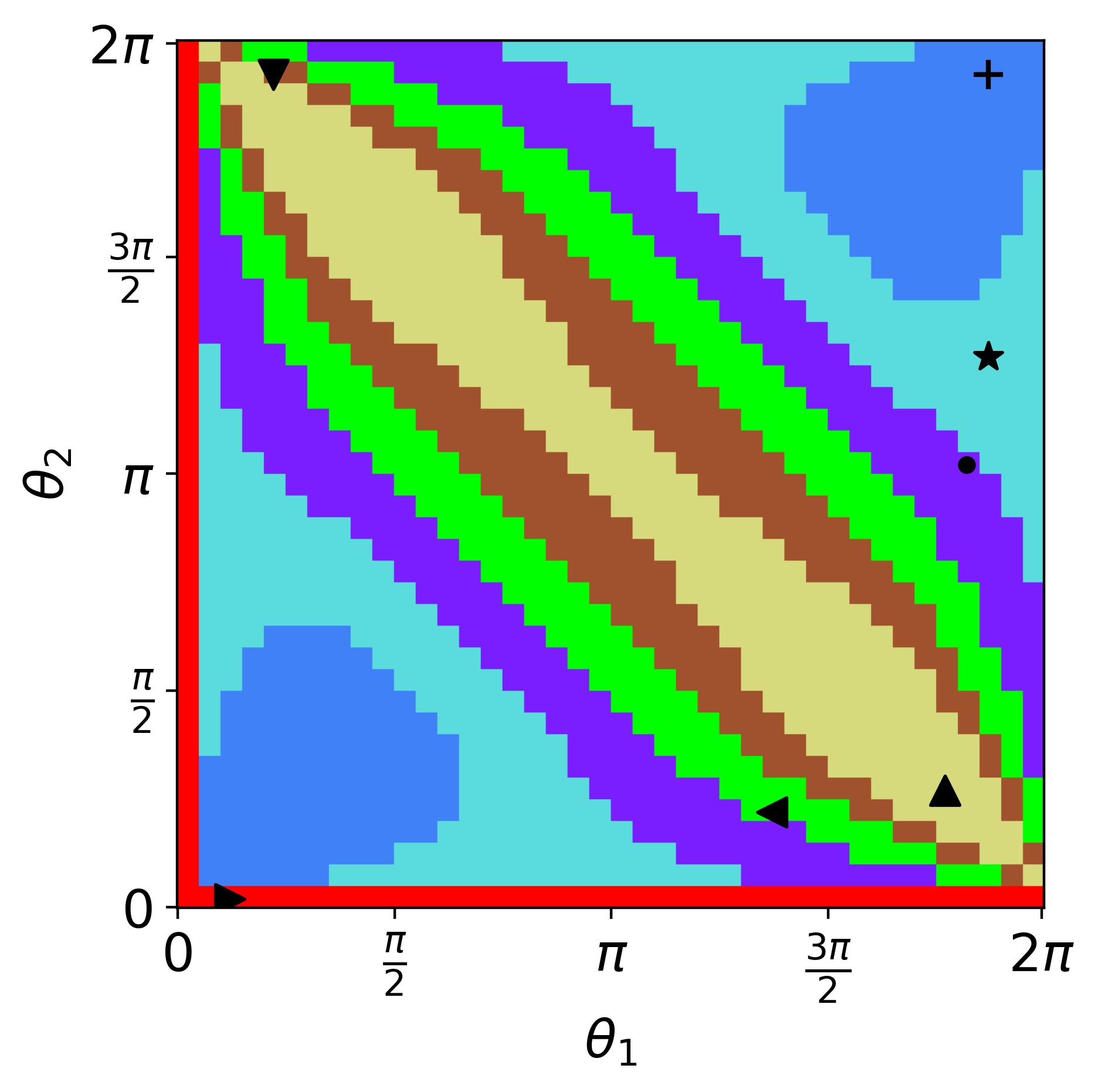}}
    \caption{(a) The clusters of PageRank in alternate equal PageRank model in Eq.~\ref{eq:AE-PR}, and the locations of the chosen typical PageRank distribution in Fig.~\ref{fig:clusters-alternate-typical}, (b) The clusters of PageRank in alternate opposite PageRank model in Eq.~\ref{eq:AO-PR}, and the locations of the chosen typical PageRank distribution in Fig.~\ref{fig:clusters-opposite-typical}, (c)The clusters of PageRank in alternate fixing PageRank model in Eq.~\ref{eq:AF-PR}, and the locations of the chosen typical PageRank distribution in Fig.~\ref{fig:clusters-fixing-typical}, (d) The clusters of PageRank in PageRank model on a trackback graph in Fig.~\ref{fig:graph}(b), and the locations of the chosen typical PageRank distribution in Fig.~\ref{fig:traceback-graph-typical}. Note: the color encodes the different cluster areas.} \label{fig:markers-alternate- trackback}
\end{figure*}

\section{Alternate quantum PageRank}
\label{sec:clusters-Modified-PageRank}
In this section, we propose the alternate quantum PageRank model and study the cluster phases therein. 
\subsection{The alternate PageRank}
The evolution operator for the alternate quantum PageRank is defined as follows,

\begin{figure*}[!ht]
    \centering
{{\subfigure(a)}\includegraphics[width=0.43\columnwidth]{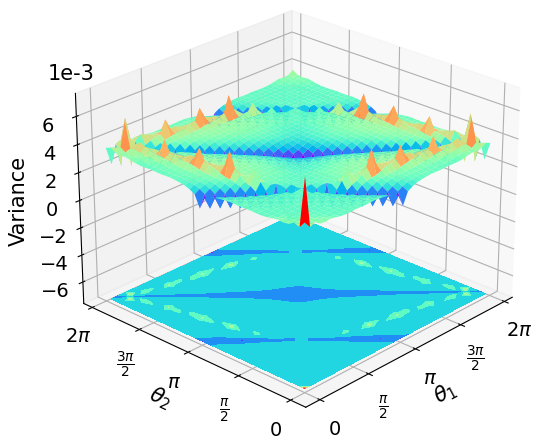}}
{{\subfigure(b)}\includegraphics[width=0.43\columnwidth]{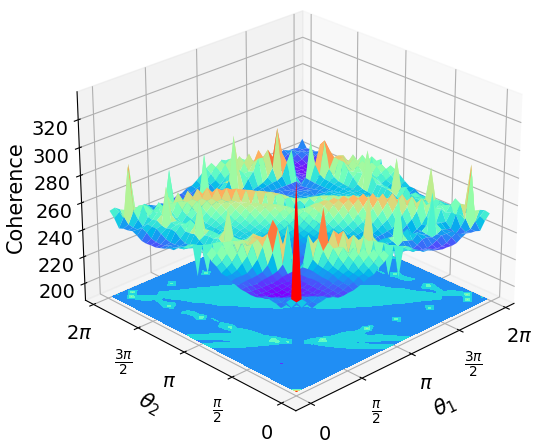}}
{{\subfigure(c)}\includegraphics[width=0.43\columnwidth]{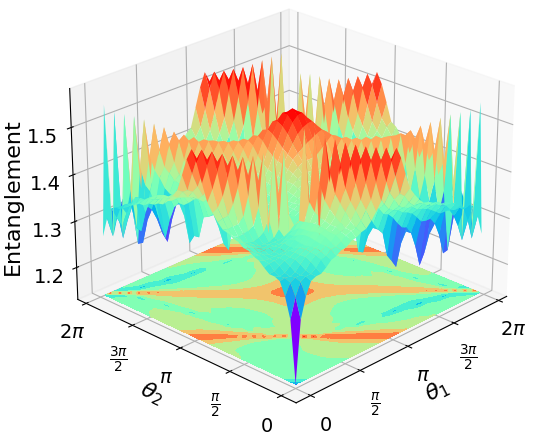}}
{{\subfigure(d)}\includegraphics[width=0.43\columnwidth]{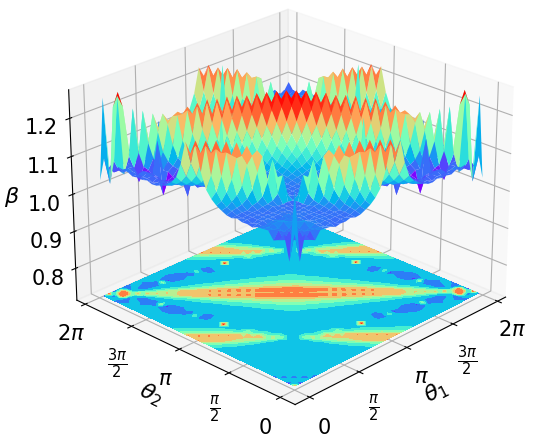}}
    \caption{(a) The distribution of the variance of the alternate opposite PageRank distributions $\protect\overrightarrow{\bf{I}}$. (b) the distribution of the alternate opposite PageRank state coherence and (c) entanglement (d) the $\beta$ power distribution      
    with respect to $\theta_1$ and $\theta_2$ in alternate opposite PageRank. Note: the color represents the value of the y axis quantity.} \label{fig:clusters-opposite}
\end{figure*}

\begin{equation}
\mathds{W}:=W(\theta'_1,\theta_1)W(\theta'_2,\theta_2),
 \label{eq:alternate-PR}
\end{equation}

where the $W$ is defined in Eq.~\ref{eq:W}, the alternate PageRank operator $\mathds{W}$ involves four independent parameters $\theta_1,\theta'_1,\theta_2,\theta'_2$, which offers rich data mining capability for PageRank. In this work, we study three specific cases involving two independent parameters by setting the dependency relation between $\theta_1, \theta'_1$ and $\theta_2,\theta'_2$, as shown below,

\textbf{Case I: the alternate equal phase PageRank}. In this case, the dependency relation is set as $\theta_1=\theta'_1$ and $\theta_2=\theta'_2$, and the resulted evolution operator is 
\begin{equation}
\mathds{W}_\text{AE}=W(\theta_1,\theta_1)W(\theta_2,\theta_2)
\label{eq:AE-PR}
\end{equation}

\textbf{Case II: the alternate opposite phase PageRank}. In this case, the dependency relation is set as $\theta_1=-\theta'_1$ and $\theta_2=-\theta'_2$, and the resulted evolution operator is 
\begin{equation}
\mathds{W}_\text{AO}=W(-\theta_1,\theta_1)W(-\theta_2,\theta_2)
\label{eq:AO-PR}
\end{equation}

\textbf{Case III: the alternate fixing phase PageRank}. In this case, we set parameter $\theta'_1=\pi,  ~\theta'_2=\pi$ and the parameters $\theta_1, \theta_2$ as free parameters. The resulting evolution operator is 
\begin{equation}
\mathds{W}_\text{AF}=W(\theta_1,\pi)W(\theta_2,\pi).
\label{eq:AF-PR}
\end{equation}

\subsection{The diversity of cluster phases in alternate quantum PageRank}
In this section, we study the cluster phases of our alternate quantum PageRank model, where the diversity of the resulting cluster phases enables superpower data mining with our alternate quantum PageRank algorithms.

\textbf{Clusters phases in the alternate equal phase PageRank.}
Governing with the evolution operator in Eq.~\ref{eq:AE-PR}, the cluster phases of PageRank variance, coherence, entanglement, and $\beta$ power in  alternate equal phase PageRank are demonstrated in  
Fig.~\ref{fig:clusters-alternate}, which are consistent with the clusters formed by the KNN clusters of PageRank distribution in Fig.~\ref{fig:markers-alternate- trackback}(a). The corresponding typical PageRank distributions in each cluster and the typical power-law relation in each cluster are shown in Fig.~\ref{fig:clusters-alternate-typical}.

The distinction of clusters in alternate equal PageRank model is the relative weights of the dominating important nodes. From the comparison of its typical PageRank distributions and the logarithmic plot, we discover that for alternate equal PageRank, all clusters break the degeneracy of residual nodes and see the same amount of the hub layers, where only inside the different layer hubs the relative importance of the nodes are different.

\begin{figure*}[!ht]
    \centering
{{\subfigure(a)}\includegraphics[width=0.8\columnwidth]{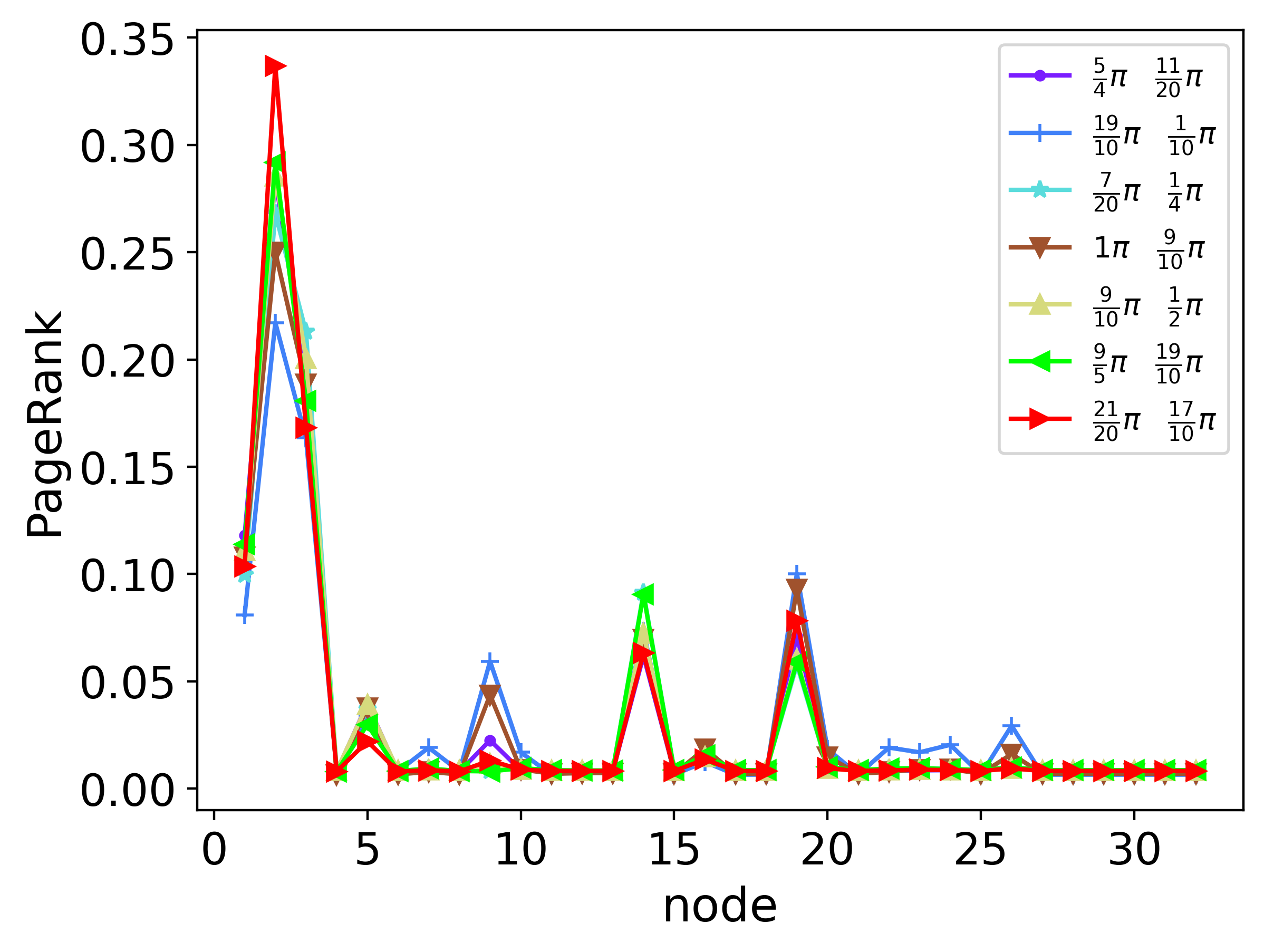}}
{{\subfigure(b)}\includegraphics[width=0.8\columnwidth]{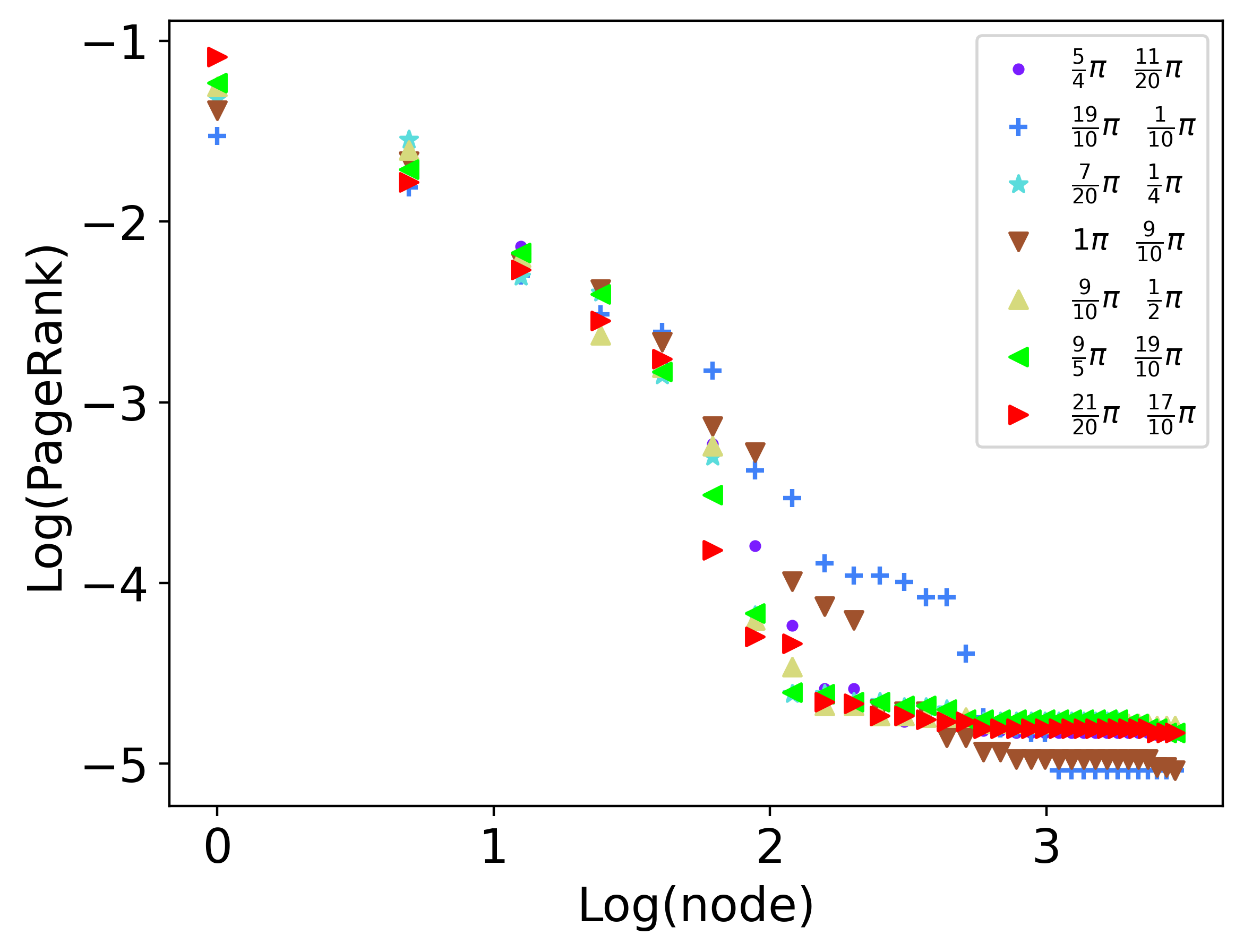}}
    \caption{(a) The comparison of typical PageRank distribution in alternate opposite quantum PageRank. (b) the comparison of the typical PageRank distribution power-law relation in alternate opposite quantum PageRank. The $(\theta_1,\theta_2)$ chosen to generate the data is noted with the same makers in Fig.~\ref{fig:markers-alternate- trackback}(b). Note: the color of the markers in (a) and (b) are consistent with the color of the clusters it belong
to in Fig.~\ref{fig:markers-alternate- trackback}(b), the values of $(\theta_1, \theta_2)$ are shown in the legend of (a).} \label{fig:clusters-opposite-typical}
\end{figure*}

\textbf{Clusters phases in the alternate opposite phase PageRank.}
Governing with the evolution operator in Eq.~\ref{eq:AO-PR}, the cluster phases of PageRank variance, coherence, entanglement and $\beta$ power in alternate opposite phase PageRank are demonstrated in  
Fig.~\ref{fig:clusters-opposite}, which are consistent with the clusters formed by the KNN clusters of PageRank distribution in Fig.~\ref{fig:markers-alternate- trackback}(b). The corresponding typical PageRank distributions in each cluster, and the typical power-law relation in each cluster are shown in Fig.~\ref{fig:clusters-opposite-typical}, where we discover for the quantum alternate opposite PageRank within anti-diagonal direction region parameters, the PageRank secondary hubs are demonstrated, the PageRank degeneracy is broken in a more obvious way, and the other clusters do not demonstrate obvious secondary hubs.

\begin{figure*}[!ht]
    \centering
{{\subfigure(a)}\includegraphics[width=0.44\columnwidth]{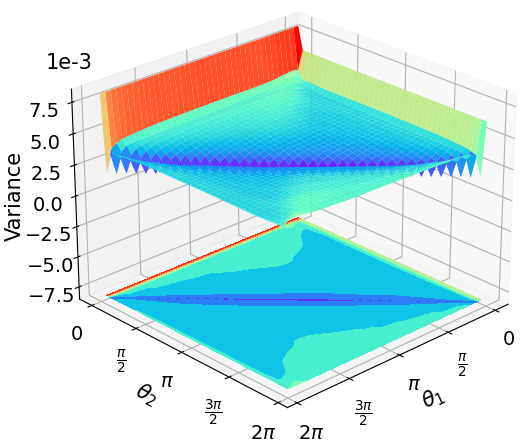}}
{{\subfigure(b)}\includegraphics[width=0.44\columnwidth]{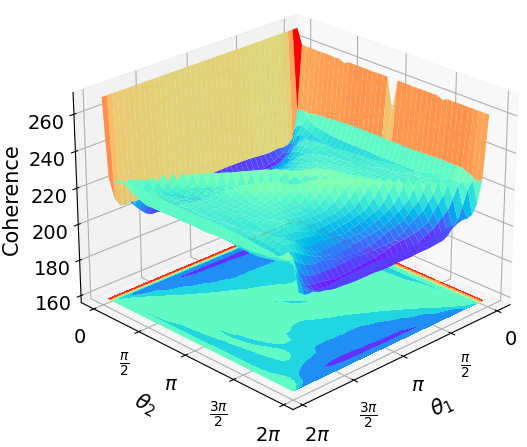}}
{{\subfigure(c)}\includegraphics[width=0.44\columnwidth]{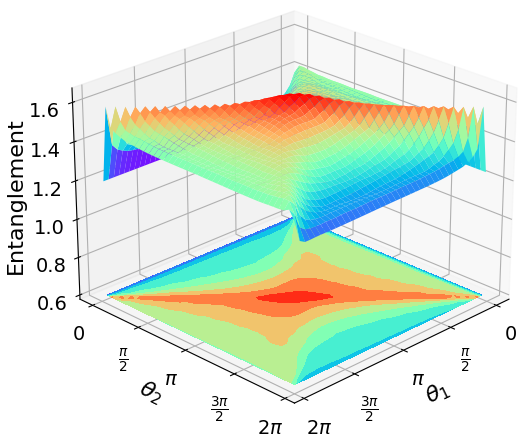}}
{{\subfigure(d)}\includegraphics[width=0.44\columnwidth]{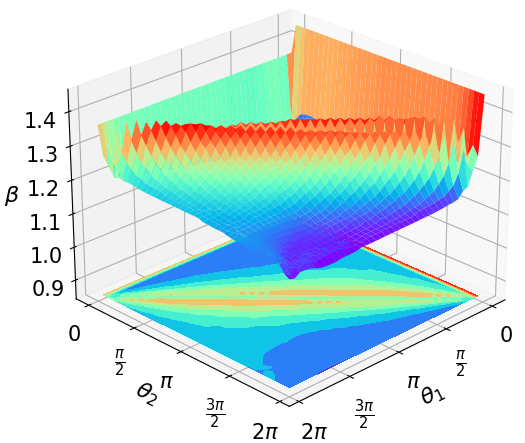}}
    \caption{(a) The distribution of the variance of the alternate fixing PageRank distributions $\protect\overrightarrow{\bf{I}}$. (b) the distribution of the alternate fixing PageRank state coherence and (c) entanglement (d) the $\beta$ power distribution      
    with respect to $\theta_1$ and $\theta_2$ in alternate fixing PageRank. Note: the color represents the value of the y axis quantity.} \label{fig:clusters-fixing}
\end{figure*}

\begin{figure*}[!ht]
    \centering
{{\subfigure(a)}\includegraphics[width=0.8\columnwidth]{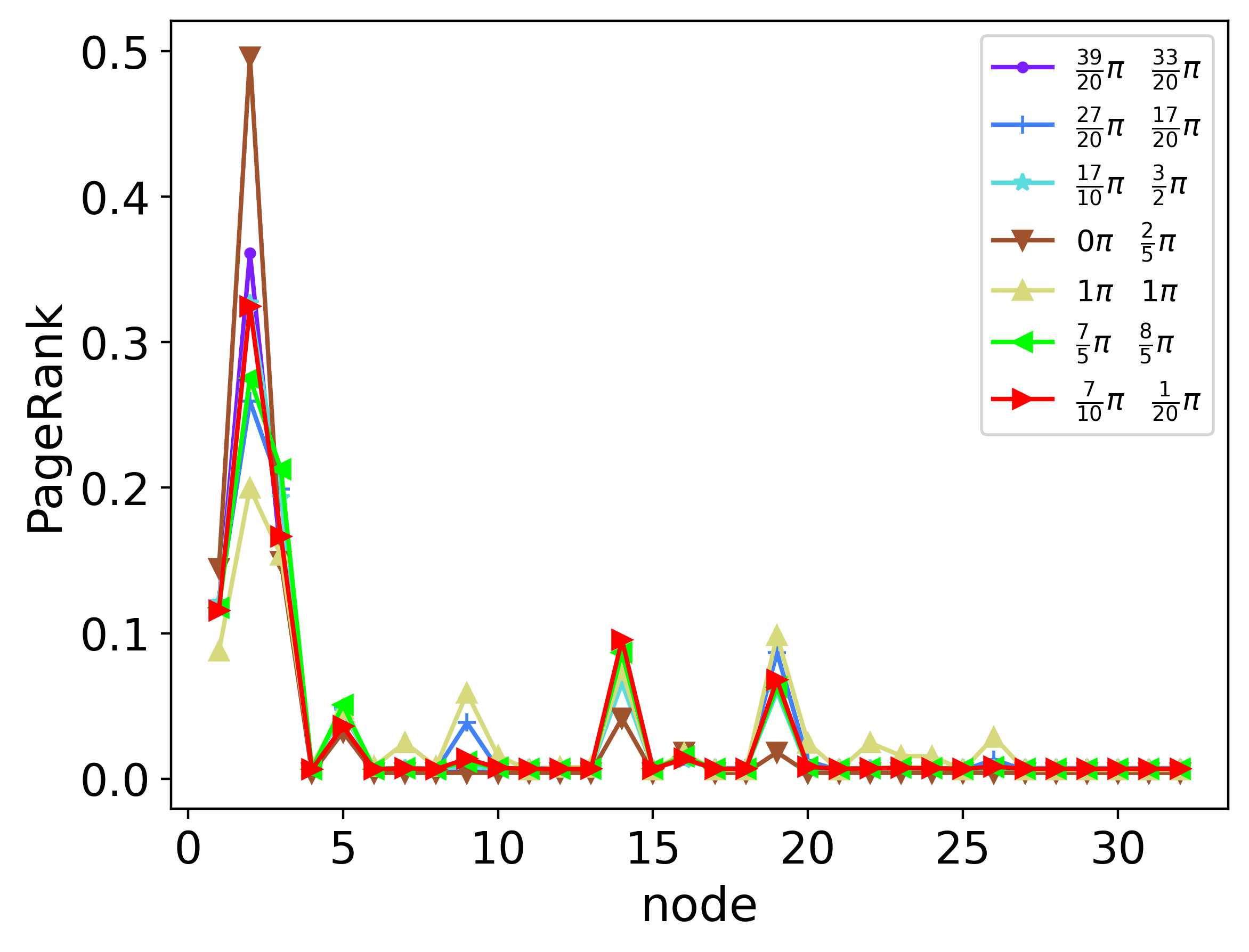}}
{{\subfigure(b)}\includegraphics[width=0.8\columnwidth]{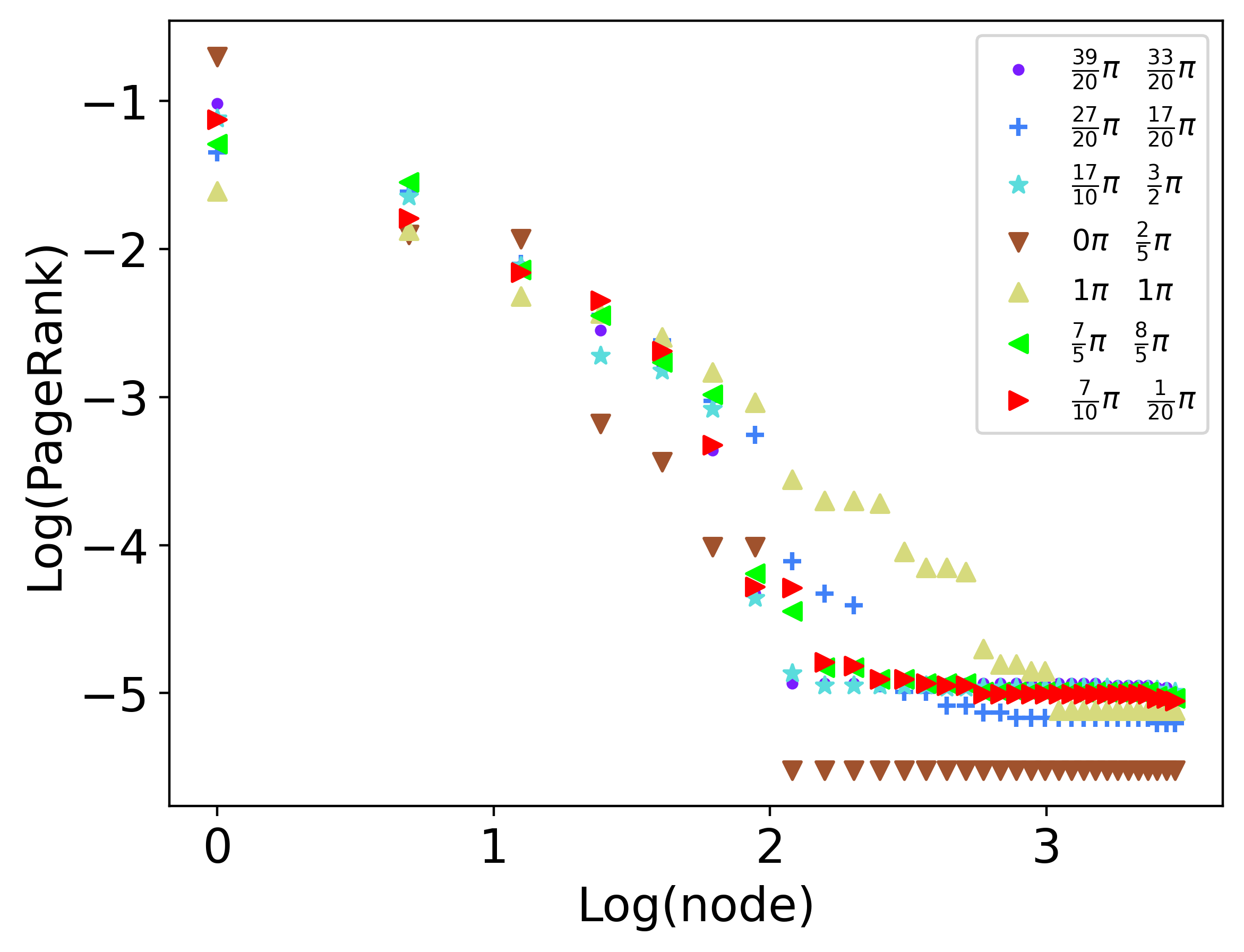}}
    \caption{(a) The comparison of typical PageRank distribution in alternate fixing quantum PageRank. (b) the comparison of the typical PageRank distribution power-law relation in alternate fixing  quantum PageRank. The $(\theta_1,\theta_2)$ chosen to generate the data is noted with the same makers in Fig.~\ref{fig:markers-alternate- trackback}(c). Note: the color of the markers in (a) and (b) are consistent with the color of the clusters it belongs
to in Fig.~\ref{fig:markers-alternate- trackback}(c), the values of $(\theta_1, \theta_2)$ are shown in the legend of (a).} \label{fig:clusters-fixing-typical}
\end{figure*}

\textbf{Clusters phases in the alternate fixing  phase PageRank.}
Governing with the evolution operator in Eq.~\ref{eq:AF-PR}, the cluster phases of PageRank variance, coherence, entanglement, and $\beta$ power in the alternate fixing phase PageRank are demonstrated in  
Fig.~\ref{fig:clusters-fixing}, which are consistent with the clusters formed by the KNN clusters of PageRank distribution in Fig.~\ref{fig:markers-alternate- trackback}(c). The corresponding typical PageRank distributions in each cluster, and the typical power-law relation in each cluster are shown in Fig.~\ref{fig:clusters-fixing-typical}, where we discover that the quantum alternate fixing PageRank with anti-diagonal direction region parameters has the largest degeneracy broken phenomenon and demonstrates the secondary and third hubs,  and the other cluster does have similar degeneracy broken behavior.

\textbf{The cluster phases in the alternate PageRank model} as shown in Fig.~\ref{fig:markers-alternate- trackback}  demonstrate the powerful data mining ability and offer various perspectives for the PageRank results. In our three specific alternate PageRank cases, we also  discover the correlations between the variance, coherence, entanglement and the $\beta$ parameter of the 
power-law relation for the PageRank in a scale-free graph. We discover the similarities between standard PageRank with APR in Eq.~\ref{eq:W} and the alternate fixing quantum PageRank in Eq.~\ref{eq:AF-PR}, where coherence, entanglement and $\beta$ are positively correlated while variance is anti-correlated with them. We also discover the similarities between standard PageRank with alternate equal quantum PageRank in Eq.~\ref{eq:AE-PR} and the alternate opposite quantum PageRank in Eq.~\ref{eq:AO-PR}, where variance, coherence, entanglement, and $\beta$ are all positively correlated. 
By the comparisons of the typical distributions of different alternate PageRank in Figs.~\ref{fig:knn-typical distribution}(b),~\ref{fig:clusters-alternate-typical}(a),~\ref{fig:clusters-opposite-typical}(a),~\ref{fig:clusters-fixing-typical}(a), we confirm that for the same scale-free graph, our model emphasizes the same set of the important nodes (nodes indexed as 1,2,3,5,7,9,10,14,19,26), and discover that the relative importance of these important nodes varies in our results, where the quantumness enable our model reveal the novel perspective interpretation of the complex networks.

\begin{figure*}[!ht]
    \centering
{{\subfigure(a)}\includegraphics[width=0.44\columnwidth]{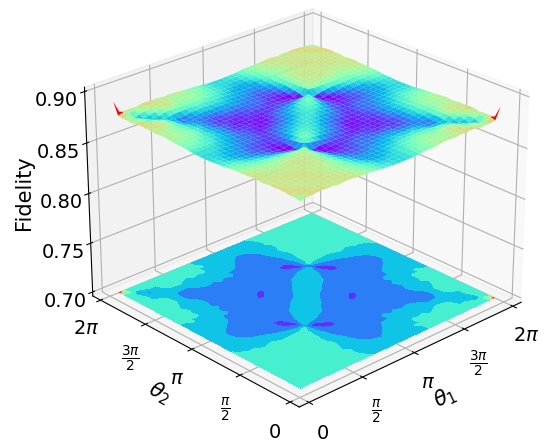}}
{{\subfigure(b)}\includegraphics[width=0.44\columnwidth]{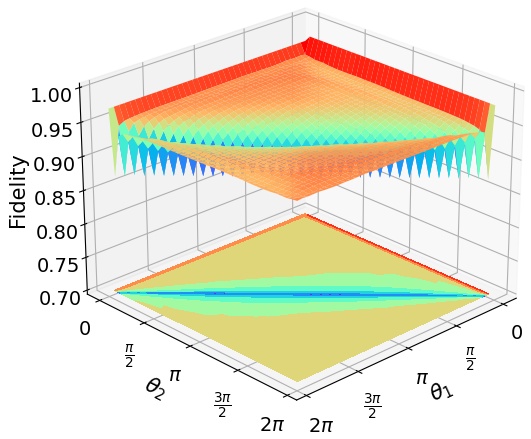}}
{{\subfigure(c)}\includegraphics[width=0.44\columnwidth]{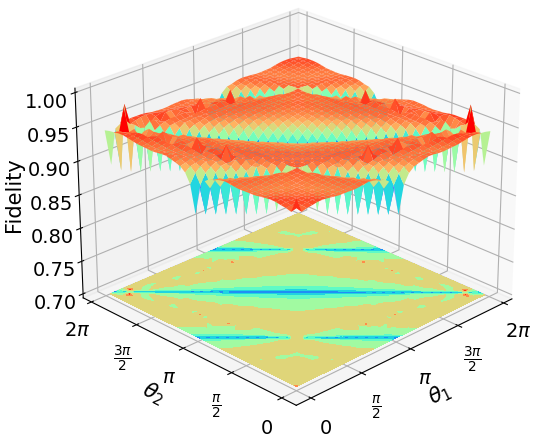}}
{{\subfigure(d)}\includegraphics[width=0.44\columnwidth]{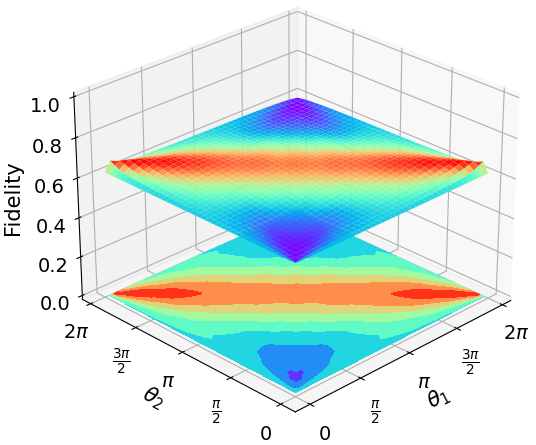}}
    \caption{(a) The distribution of fidelity between the quantum PageRank distributions in ($\theta_1,\theta_2$) space from  alternate equal PageRank model in Eq.~\ref{eq:AE-PR}, and the classical PageRank distribution, (b) The distribution of fidelity between the quantum PageRank distributions in ($\theta_1,\theta_2$) space from alternate opposite PageRank model in Eq.~\ref{eq:AO-PR}, and the classical PageRank distribution, (c)The distribution of fidelity between the quantum PageRank distributions in ($\theta_1,\theta_2$) space from alternate fixing PageRank model in Eq.~\ref{eq:AF-PR}, and the classical PageRank distribution, (d) The distribution of quantum PageRank fidelity between the quantum PageRank distributions in ($\theta_1,\theta_2$) space from  quantum  PageRank model in Eq.~\ref{eq:AE-PR}  on a trackback graph in Fig.~\ref{fig:graph}(b), and the corresponding classical PageRank distribution. Note: the color represents the value of the fidelity quantity. \label{fig:similarities-four}}
    
\end{figure*}

\begin{figure*}[t]
    \centering
{{\subfigure(a)}\includegraphics[width=0.44\columnwidth]{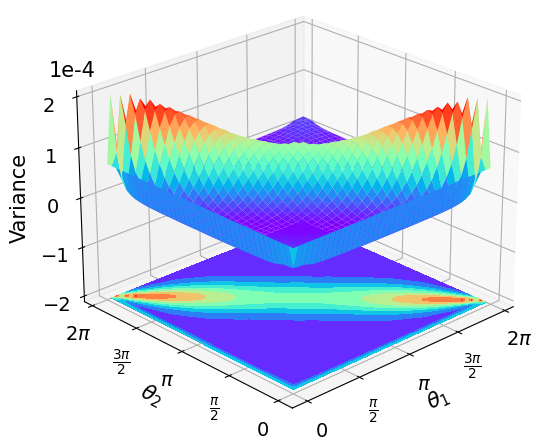}}
{{\subfigure(b)}\includegraphics[width=0.44\columnwidth]{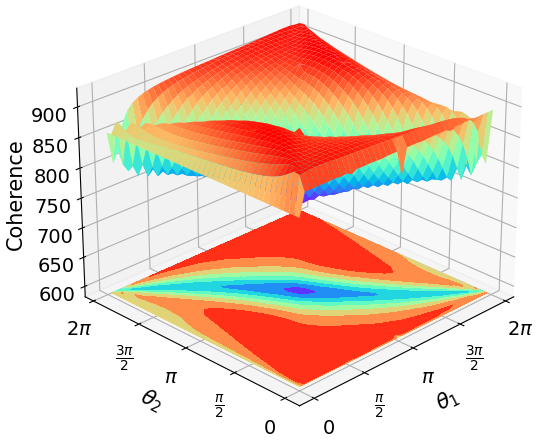}}
{{\subfigure(c)}\includegraphics[width=0.44\columnwidth]{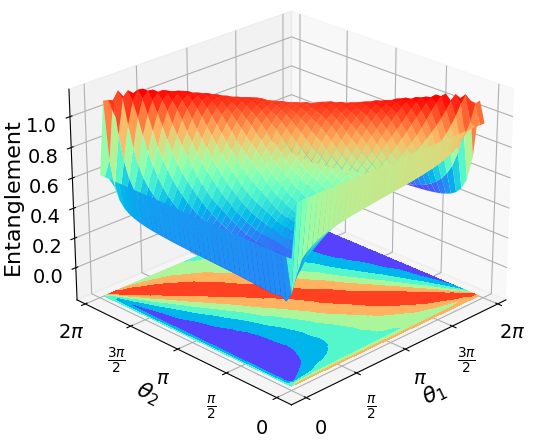}}
{{\subfigure(d)}\includegraphics[width=0.44\columnwidth]{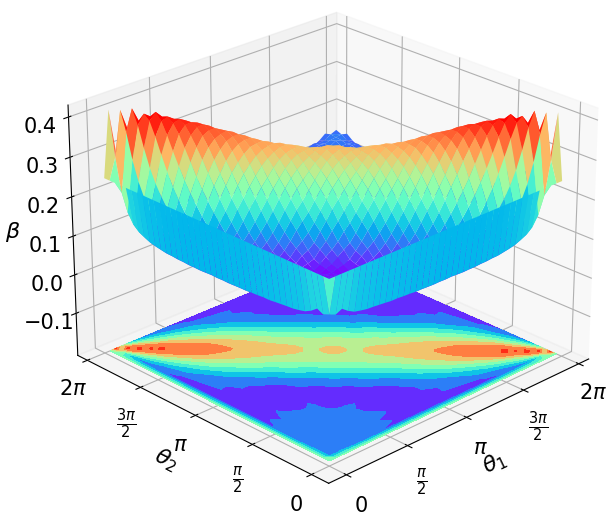}}
    \caption{The clusters of the PageRank on the trackback graph in Fig.~\ref{fig:graph}(b): (a) The distribution of the variance of the PageRank distributions $\protect\overrightarrow{\bf{I}}$. (b) the distribution of  PageRank state coherence and (c) entanglement (d) the $\beta$ power distribution      
    with respect to $\theta_1$ and $\theta_2$ in alternate fixing PageRank. Note: the color represents the value of the y axis quantity.} \label{fig:clusters-traceback}
\end{figure*}

The quantumness in alternate quantum PageRank can be observed from the quantum PageRank fidelity distribution in the $\{\theta_1,\theta_2\}$ space between the alternate quantum PageRank distributions and the corresponding classical PageRank distribution, as seen in Fig.~\ref{fig:similarities-four} and Fig.~\ref{fig:markers-alternate- trackback}, where for all our alternate PageRank schemes, the quantum PageRank fidelity distribution demonstrate similar cluster patterns. The positive correlation between the quantum PageRank fidelity and the variance of the PageRank probability distribution, the anti-correlation between the quantum PageRank fidelity with the classical PageRank, and coherence, entanglement,  the $\beta$ coefficient of the quantum PageRank still hold for our alternate PageRank schemes, same as our standard quantum PageRank with APR. 
Same as previously discussed, for alternate quantum PageRank schemes the quantumness in the schemes is anti-correlated with the variance of the PageRank, and the fidelity with the classical PageRank, but positive-correlated with the coherence, entanglement, and the $\beta$.

\section{The PageRank of a scale-free graph's trackback graph}
In this section, we study the PageRank on the trackback graph of a scale-free graph as shown in Fig.~\ref{fig:graph}(b), which is the reverse graph of the 32-node scale-free graph in Fig.~\ref{fig:graph}(a) and have the reversed directed edges. 

\begin{figure*}[ht]
    \centering
{{\subfigure(a)}\includegraphics[width=0.8\columnwidth]{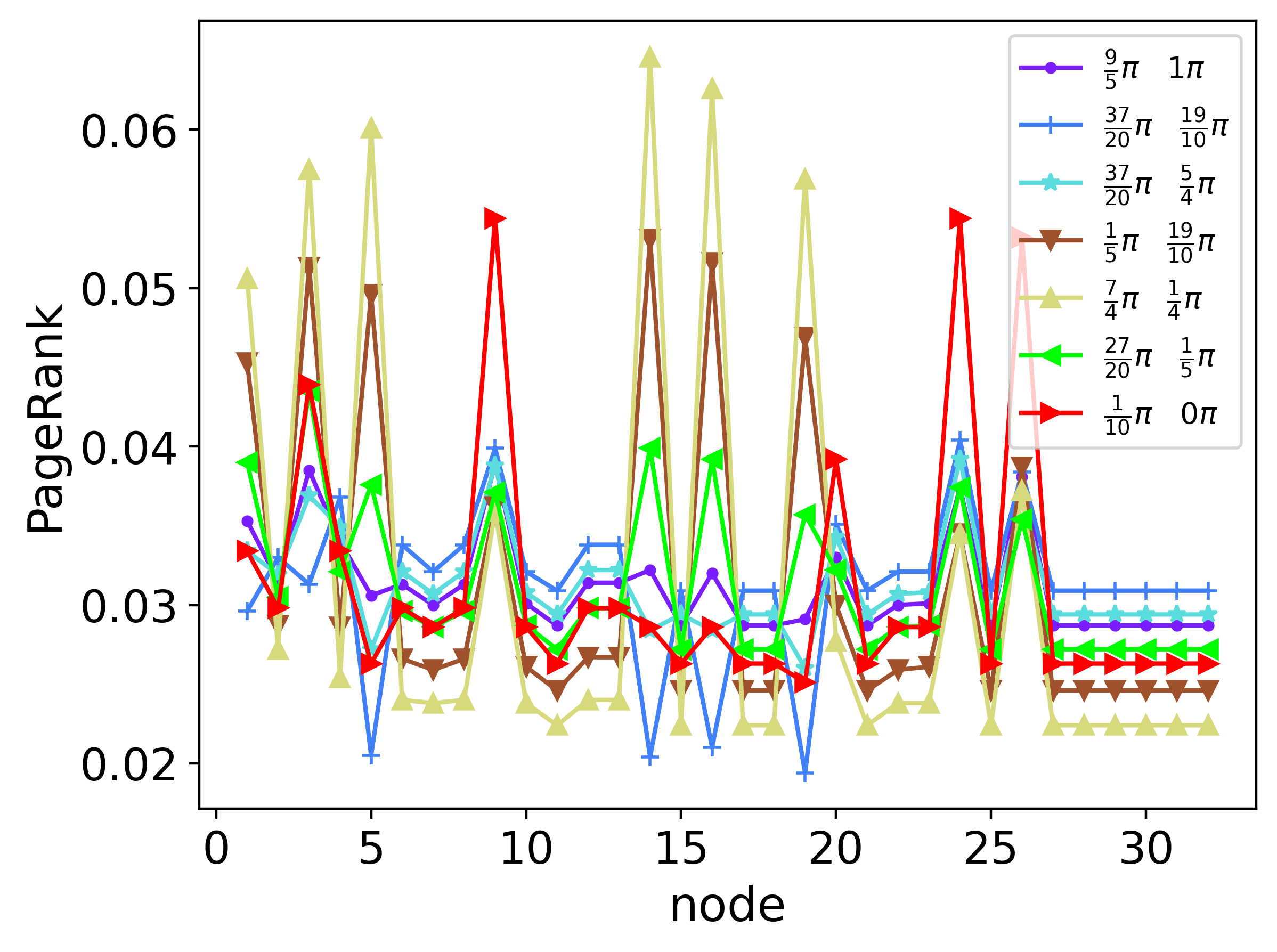}}
{{\subfigure(b)}\includegraphics[width=0.8\columnwidth]{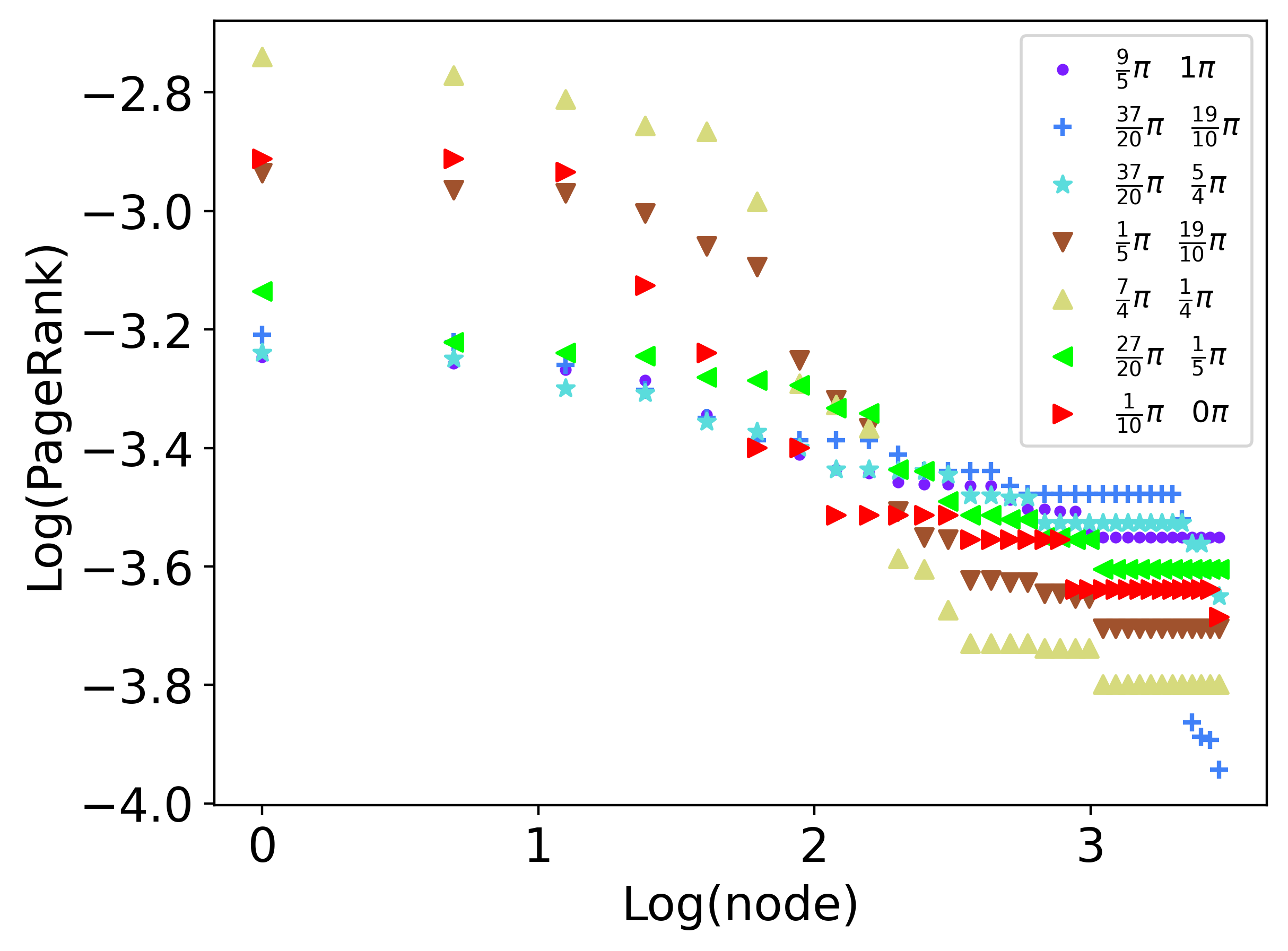}}
    \caption{(a) The comparison of typical PageRank distribution in quantum PageRank on the trackback graph. (b) the comparison of the typical PageRank distribution power-law relation in quantum PageRank on the trackback graph. The $(\theta_1,\theta_2)$ chosen to generate the data is noted with the same makers in Fig.~\ref{fig:markers-alternate- trackback}(d). Note: the color of the markers in (a) and (b) are consistent with the color of the clusters it belong
to in Fig.~\ref{fig:markers-alternate- trackback}(d), the values of $(\theta_1, \theta_2)$ are shown in the legend of (a).} \label{fig:traceback-graph-typical}
\end{figure*}

For the PageRank in a trackback graph mentioned above, the distribution of quantum PageRank fidelity in
Fig.~\ref{fig:similarities-four}(c) and the distributions of the variance of the PageRank distribution, coherence and entanglement of the PageRank states, and the $\beta$ parameter of the power-law relation  shown in Fig.~\ref{fig:clusters-traceback}  are consistent with the clusters formed by the KNN clusters of PageRank distribution in Fig.~\ref{fig:markers-alternate- trackback}(d). The typical PageRank distributions in each cluster and the power-law relations are shown in Fig.~\ref{fig:traceback-graph-typical}, where we see the $\beta$ curve does not have an obvious linear behavior since the graph is not a scale-free graph anymore. But we do observe the multi-degeneracy phenomena which indicate the multi-importance node hubs in the trackback graph. Different clusters evaluate the nodes' importance differently and signal different numbers of degenerated hubs, which offers a potential tool for data analysis in complex networks.

For our quantum PageRank with APR on the trackback graph, we discover a positive correlation between the quantum PageRank fidelity,  and the coherence of the quantum PageRank, an anti-correlation with the 
variance of the PageRank probability distribution, entanglement, and the $\beta$ coefficient of the quantum PageRank. We conjecture that, 
the coherence captured here is more related to the classical correlation instead of a quantum correlation.  Therefore, it is positively correlated with the quantum PageRank fidelity. The variance, entanglement, and the $\beta$ coefficient truly reflect the quantumness.
The typical PageRank Probability distribution in each cluster and the typical $\beta$ power distributions are shown in Fig.~\ref{fig:traceback-graph-typical}. We discover that all the typical PageRank Probability distributions emphasize different important nodes set (nodes indexed as 1,3,5,9,14,16,19,20,24,26) from the original scale-free graph, which provides new insights for the complex network tracking.

The trackback graph plays an important role in the field of network attack and defense. The structural features of the traceback graph are closely related to network attack detection algorithms and prevention strategies. Identifying key nodes and structural holes in the network and implementing targeted defense measures can optimize the network's defense layout, which is the purpose our PageRank models serve.

By the comparison of the cluster patterns of our four quantum PageRank schemes, we discover that the standard quantum PageRank model and alternate equal PageRank model enable a diagonal direction originated pattern, while the alternate opposite PageRank, alternate fixing PageRank
model, and PageRank model on a trackback graph demonstrate an anti-diagonal direction originated patterns. With the ability to support different clusters, we can synthesize quantum systems under complex environmental constraints and enable the practical application of quantum PageRanks.

We investigate our schemes on two extra scale-free graphs with 16 and 20 nodes separately, and the results are shown in the supplementary file. We discover the Quantum PageRank KNN clusters,  the quantum PageRank fidelity with the corresponding classical PageRanks,  variance, coherence, entanglement, and the $\beta$ parameter are all correlated which is consistent with our main results but whether positively correlated or anti-correlated varies as the graph is different.

\section{Conclusions} \label{sec:conclusions}
In this work, we demonstrate the quantum versatility in PageRank by studying the cluster phases in PageRank distributions, the quantum PageRank fidelity with the corresponding classical PageRank, the variance, coherence, and entanglement therein, and the $\beta$ parameter in its power-law fitting. We propose an alternate PageRank model that offers various quantum-enabled perspectives for  PageRank. We discover the correlations between the variance, the quantum PageRank fidelity with the corresponding classical PageRank, and the $\beta$ parameter in its power-law fitting, coherence, and entanglement. The consistency of the cluster phases and phase boundaries provides a novel perspective for the study of quantum PageRank and sheds light on quantum data mining.

\section*{
CRediT authorship contribution statement}
Wei-Wei Zhang: Conceptualization, Methodology, Writing original draft, supervision.  
Zheping Wu: Methodology, Numerical Simulation, Writing, Review and
Editing. 
Wei Zhao, Qingbing Ji, Hengyue Jia, Wei Pan: Investigation, Writing, Review and Editing, Validation. Haobing Shi: Visualization, Writing and
Review and Editing, Supervision.

\section*{
Declaration of competing interest}
The authors declare that they have no known competing financial interests or personal relationships that could have appeared to 
influence the work reported in this paper.

\section*{
Data availability}

All data included in this study are available upon request by contacting authors.

\section*{Acknowledgements} \label{sec:acknowledgements}
  We acknowledge support from the National Natural Science Foundation of China (Grant No.  12104101) and the Fundamental Research Funds for the Central Universities,Stability Program of National Key Laboratory of Security Communication (2023), the Major Research Project of National Natural Science Foundation of China under Grant 92267110, the Joint Funds of the National Natural Science Foundation of China (Grant No. U22B2025) and the Key Research and Development Program of Shaanxi  (Grant No. 2023-GHZD-42).

$^\dagger$ W.Z and Z.W contribute equally.

\end{document}